\documentclass[aps,prb,10pt,twocolumn,superscriptaddress]{revtex4-2}

\usepackage[utf8]{inputenc}
\usepackage{graphicx}
\usepackage{amssymb,amsmath,amsfonts}
\usepackage{color}
\usepackage[bookmarks=false]{hyperref}

\DeclareMathOperator{\sgn}{sgn}

\begin{document}


\title{Superconducting phases and the second Josephson harmonic in tunnel junctions between diffusive superconductors}
\author{A.~S.\ Osin}
\affiliation{Moscow Institute of Physics and Technology, 141700 Dolgoprudny, Russia}
\affiliation{Skolkovo Institute of Science and Technology, 143026 Moscow, Russia}
\affiliation{L.~D.\ Landau Institute for Theoretical Physics RAS, 142432 Chernogolovka, Russia}

\author{Ya.~V.\ Fominov}
\affiliation{L.~D.\ Landau Institute for Theoretical Physics RAS, 142432 Chernogolovka, Russia}
\affiliation{Laboratory for Condensed Matter Physics, HSE University, 101000 Moscow, Russia}

\date{15 August 2021}

\begin{abstract}
We consider a planar SIS-type Josephson junction between diffusive superconductors (S) through an insulating tunnel interface (I). We construct fully self-consistent perturbation theory with respect to the interface conductance. As a result, we find correction to the first Josephson harmonic and calculate the second Josephson harmonic. At arbitrary temperatures,
we correct previous results for the nonsinusoidal current-phase relation in Josephson tunnel junctions, which were obtained with the help of conjectured form of solution.
Our perturbation theory also describes the difference between the phases of the order parameter and of the anomalous Green functions.
\end{abstract}

\maketitle

\tableofcontents

\section{Introduction}
\label{sec:one}

One of the key characteristics of a superconductor is the complex-valued order parameter $\Delta(\mathbf{r})$, which is parametrized by its absolute value and phase $\varphi(\mathbf{r})$ \cite{TinkhamBook}.
Both parameters are essential for describing current-carrying states of superconductors.
While the absolute value of the order parameter determines the density of the superconducting condensate, the phase gradient is related to the condensate velocity. At the same time, more detailed spectral (i.e., energy-resolved) information about superconductivity in a system is contained in the anomalous Green function $F(\mathbf{r},\omega)$ (here $\omega$ is the Matsubara frequency), with its own absolute value and phase $\chi(\mathbf{r},\omega)$. The anomalous Green functions and the order parameter (related by the self-consistency equation) fully describe superconductivity inside an equilibrium system \cite{AGDBookEng}.

The Josephson effect is a prominent example of the physical role of the superconducting phases \cite{TinkhamBook}. The simplest Josephson system is a planar SIS-type junction (superconductors S separated by an insulating barrier I).
All characteristics of the system depend on a single coordinate $x$ (normal to the plain interface).
Fully self-consistent treatment of the Josephson effect in the SIS junction requires taking into account difference between the phases of the order parameter and the anomalous Green function,  $\varphi(x)\neq\chi(x,\omega)$.
Although difference between the two phases is a well-known fact (which is already evident from frequency, or energy, dependence of $\chi$ while $\varphi$ depends only on coordinate)
\cite{Zaikin1981Eng,Stoof1996.PhysRevB.53.14496,Belzig1999}, it has been taken into account in actual calculations mainly numerically \cite{Golubov2002Eng}.

At the same time, the SIS junction is the fundamental system for which the Josephson effect was originally predicted \cite{Josephson1962,Ambegaokar1963}, and it has been considered in many various limiting cases. In the main order with respect to the interface conductance, the Josephson current proportional to the sine of the order-parameter phase difference between the banks arises, $J\propto \sin\delta\varphi$. Next orders with respect to the interface conductance take into account additional effects such as pair-breaking due to current and the proximity effect between the banks (suppression of the order parameter near the interface) \cite{Likharev1979.RevModPhys.51.101,Golubov2004review}. These effects influence basic characteristics of the Josephson current such as the value of the critical current and the current-phase relation in SIS and more complicated types of Josephson junction (including SNS junctions with normal metal N as a weak link) \cite{Ivanov1981Eng,Kupriyanov1982Eng,Zubkov1983Eng}. As a result, the current-phase relation $J(\delta\varphi)$ can deviate from the simple sinusoidal form \cite{Ivanov1981Eng,Zubkov1983Eng}.

Anharmonic (nonsinusoidal) Josephson current is also possible in the case of pair-breaking interfaces \cite{Barash2012.PhysRevB.85.100503,Barash2014Eng}. SIS junctions with arbitrary interface transparency have been considered in the limit of temperature close to the critical one \cite{Sols1994.PhysRevB.49.15913,Pastukh2017}.

In this paper, we consider the SIS Josephson junction between diffusive superconductors at arbitrary temperature $T$. We consider the tunneling limit but focus on deviations from the sinusoidal current-phase relation due to small but finite conductance of the interface.
We develop fully self-consistent perturbation theory taking into account difference between the phases $\varphi(x)$ and $\chi(x,\omega)$. We carry out perturbation theory with respect to the ratio of the interface conductance to the conductance of the superconductor on the coherence length, and this parameter is small in the tunneling limit.
As a result, we find the second harmonic of the Josephson relation, i.e., contribution to $J(\delta\varphi)$ of the form $\sin 2\delta\varphi$.
The first harmonic $\sin\delta\varphi$ arises in the first order of the perturbation theory, while the second harmonic $\sin2\delta\varphi$ comes from the second order, thus being small in comparison with the first one. In the limit $T\to T_c$, we reproduce the results by Kupriyanov \cite{Kupriyanov1992}.
At arbitrary temperatures, we revisit the results by Golubov and Kupriyanov \cite{GK2005}. In Ref.\ \cite{GK2005}, the authors employed a conjectured form of solution, which turns out to be only qualitatively correct. As a result, they obtained parametrically correct answer but with wrong numerical coefficients. Our perturbation theory reproduces their parametrical results and provides exact numerical coefficients.

We also discuss quantitative difference between the phase of the order parameter and the phases of the anomalous Green functions that follows from our perturbation theory.

The paper is organized as follows:
In Sec.~\ref{sec:two}, we formulate our model and equations.
In Sec.~\ref{sec:three}, we develop our perturbation theory with respect to the interface conductance and calculate the Josephson current.
In Sec.~\ref{sec:four}, we develop our perturbation theory further and discuss difference between the phase of the order parameter and the phases of the anomalous Green functions.
In Sec.~\ref{sec:Discussion}, we discuss the applicability conditions of our perturbation theory and the role of self-consistency.
In Sec.~\ref{sec:Conclusions}, we present our conclusions.
Finally, some details of calculations are presented in the Appendices.

Throughout the paper, we employ the units with $\hbar=k_B=1$.

\section{Formalism}
\label{sec:two}

\subsection{Basic equations}

SIS-type junction is a system of two superconductors separated by a thin insulating layer.
\begin{figure}[t]
\centerline{\includegraphics[width=\columnwidth]{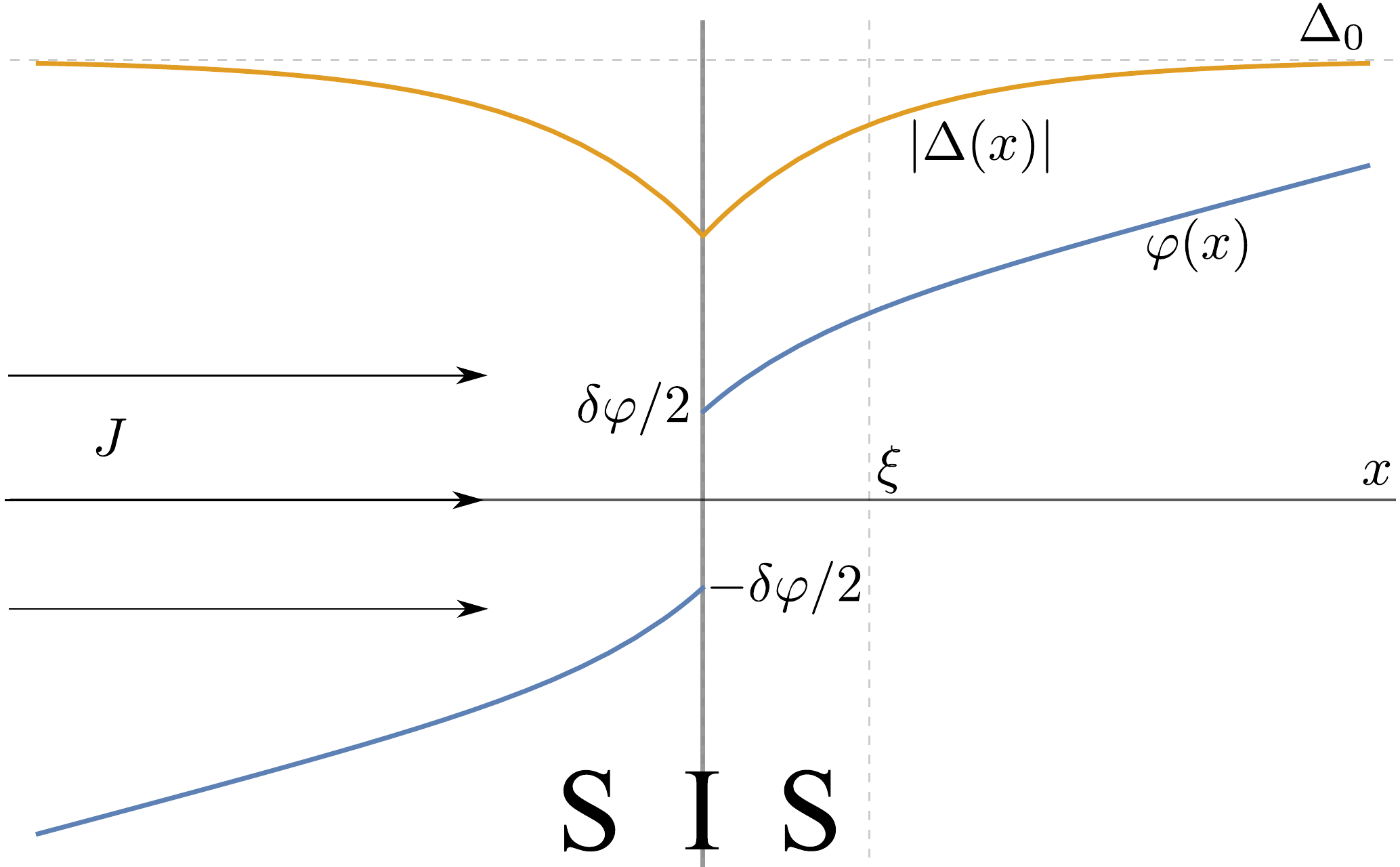}}
\caption{Sketch of a planar SIS junction. Two superconductors (S) are separated by a thin insulating layer (I). At the interface, the order-parameter phase $\varphi$ is discontinuous. The absolute value of the order parameter $|\Delta|$ is suppressed at $x=0$, while at the bulk it reaches the value $\Delta_0$. Both plots are schematic. Due to the spatial symmetry of the problem, the coordinate dependence $|\Delta(x)|$ is even while $\varphi(x)$ can be chosen odd. We parametrize the Josephson current $J$ in the junction by the phase difference $\delta\varphi\equiv\varphi(+0)-\varphi(-0)$ at the interface.}
\label{FigureSIS}
\end{figure}

In the diffusive, or so-called dirty, limit, when the superconducting coherence length $\xi$ is much larger than the mean free path $l$, superconductors can be described by the Usadel equations \cite{Usadel1970}, which are written for the isotropic (in the momentum space) parts of the quasiclassical Green functions, normal function $G(\mathbf{r},\omega)$ and anomalous function $F(\mathbf{r},\omega)$.

In the planar SIS junction all characteristics depend only on the $x$ coordinate. We can eliminate the vector potential by a gauge transformation, so that all current-carrying properties of the system are encoded in the phase gradients. In  the angular parametrization \cite{Stoof1996.PhysRevB.53.14496,Belzig1999}, $G(x,\omega)=\cos\theta(x,\omega)$, $F(x,\omega)=\sin\theta(x,\omega) e^{i\chi(x,\omega)}$, the Usadel equations take the form (see Appendix \ref{Appendix A} for more details)
\begin{align}
    &\frac{D}{2}\frac{d^{2}\theta}{dx^{2}}-\omega\sin\theta+\left|\Delta\right|\cos\left(\chi-\varphi\right)\cos\theta
    \notag\\
    &\hphantom{\frac{D}{2}\frac{d^{2}\theta}{dx^{2}}+\left|\Delta\right|}-\frac{D}{2} \left(\frac{d\chi}{dx}\right)^{2} \sin\theta\cos\theta=0,\label{Diffusion}\\
    &\frac{D}{2}\frac{d}{dx}\left( \frac{d\chi}{dx} \sin^{2}\theta\right)=\left|\Delta\right|\sin\left(\chi-\varphi\right)\sin\theta,\label{Continuity}
\end{align}
where $\Delta(x)=|\Delta(x)|e^{i\varphi(x)}$ is the order parameter, $\omega=\pi T\left(2n+1\right)$ is the Matsubara frequency (at temperature $T$), $D=v_F l/3$ is the diffusion constant, and $v_F$ is the Fermi velocity.

The Usadel equations must be accompanied by the self-consistency equation
\begin{align}
    \left|\Delta(x)\right|=\pi\lambda T\sum_{\left|\omega\right|<\omega_{D}}e^{i(\chi(x,\omega)-\varphi(x))}\sin\theta(x,\omega),{\label{Self-consistency}}
\end{align}
where $\lambda$ is the BCS coupling constant and $\omega_D$ is the Debye frequency of the superconducting material.

We consider the tunneling limit in which the Kupriyanov-Lukichev (KL) boundary conditions apply \cite{KL1987}. In the angular parametrization (see Appendix \ref{Appendix A}), they can be written as
\begin{gather}
        \frac{d\theta(\pm0,\omega)}{dx}=\pm\frac{g_N}{2\sigma}\sin2\theta(0,\omega)\left[1-\cos\delta\chi(\omega)\right]\label{Theta boundary},\\
        \frac{d\chi(\pm0,\omega)}{dx}=\frac{g_N}{\sigma}\sin\delta\chi(\omega){\label{Velocity boundary}},\\
        \delta\chi(\omega)\equiv\chi(+0,\omega)-\chi(-0,\omega){\label{Phase shift definition}},
\end{gather}
where $\sigma$ is the normal-state conductivity of the superconductor material, $g_N=G_N/S$ is the conductance of the interface per unit area, $S$ is the interface area, and $G_N$ is the interface conductance (normal-state conductance of the weak link).

In addition to the boundary conditions at $x=0$, we have to take into account that far from the interface the superconductors are in the bulk current-carrying state: the density and the velocity of the condensate become constant (do not depend on $x$),
\begin{gather}
\theta(x\to\pm\infty,\omega)=\text{const},\\
\frac{d\chi(x\to\pm\infty,\omega)}{dx}=\frac{d\varphi(x\to\pm\infty)}{dx}=\text{const}.
\end{gather}

The current $J$ can be found after solving the set of Eqs.\ (\ref{Diffusion})--(\ref{Self-consistency}), (\ref{Theta boundary}), and (\ref{Velocity boundary}) with the use of
\begin{equation} \label{Current relation}
        J=2\pi e \nu_{0}SDT \sum_{\omega} \frac{d\chi(x,\omega)}{dx} \sin^{2}\theta(x,\omega) ,
\end{equation}
where $e$ is the charge of electron and $\nu_0$ is the density of states at the Fermi level in the normal state.

The self-consistency equation guaranties the current conservation \cite{Svidzinsky1982,Furusaki1991,Bagwell1994.PhysRevB.49.6841,Sols1994.PhysRevB.49.15913}. To see that, one can take the imaginary part of  Eq.\ (\ref{Self-consistency}), which yields
\begin{equation}
    0=\sum_{\left|\omega\right|<\omega_{D}}\sin\left[\chi(x,\omega)-\varphi(x)\right]\sin\theta(x,\omega).\label{ImDelta0}
\end{equation}
Then, summing Eq.\ (\ref{Continuity}) over $\omega$ and applying Eqs.\ (\ref{Current relation}) and (\ref{ImDelta0}), we find
\begin{equation}
    \frac{dJ}{dx}\propto\sum_{|\omega|<\omega_{D}}\sin\left[\chi(x,\omega)-\varphi(x)\right]\sin\theta(x,\omega)=0,
\end{equation}
which means that $J=\text{const}$.
Therefore, the current can be found at the $x=0$ interface with the help of the KL boundary condition (\ref{Velocity boundary}).

At the same time, according to Eq.\ (\ref{Current relation}), the current can be written as a sum of spectral currents, $J= \sum_\omega J_\omega$.
The ``continuity'' equation (\ref{Continuity}) implies that in the case of $\chi\neq \varphi$, the spectral currents are not conserved, $d J_\omega/dx\neq 0$.
The distribution of the total (conserved) current between the Matsubara frequencies then varies as a function of coordinate.

\subsection{Tunneling limit}

The self-consistent Usadel equations cannot be solved analytically for arbitrary transparencies, but in some limiting cases this can be done approximately. In this paper, we solve the Usadel equations by the perturbation theory with respect to the interface conductance.

In a superconductor, the natural energy scale is the bulk temperature-dependent value of the order parameter $\Delta_0(T)$.
It determines the coherence length, which can be written (in the diffusive limit) as
\begin{gather} \label{Coherence length definition}
    \xi(T)=\sqrt{D/2\Delta_{0}(T)}.
\end{gather}
This characteristic length follows from the Usadel equations; however, it turns out to be indeed the relevant spatial scale on which the superconducting properties vary, only at temperatures not too close to the superconducting critical temperature $T_c$. In the vicinity of $T_c$, the full set of the Usadel equations can be reduces to the Ginzburg-Landau (GL) equation written for the order parameter only. In the course of this reduction, the Matsubara summation in the self-consistency equation generates a different coherence length, which can be written as
\begin{equation} \label{xiGL}
    \xi_\mathrm{GL}(T)=\sqrt{\pi D / 8 (T_c-T)}.
\end{equation}
Although this GL coherence length arises when considering the $T\to T_c$ limit, the resulting expression can be used at any $T$. From this point of view, we can say that at $T$ not too close to $T_c$, the GL coherence length (\ref{xiGL}) is of the same order as the Usadel coherence length (\ref{Coherence length definition})
\footnote{In particular, at $T=0$ we have $\xi(0)/\xi_\mathrm{GL}(0) = 2 e^{C/2}/\pi \approx 0.85$, where $C\approx 0.577$ is Euler's constant}.
However, at $T\to T_c$, they are parametrically different since $\Delta_0(T) \propto \sqrt{T_c-T}$, and $\xi_\mathrm{GL}$ is turns out to be the actual scale for $\Delta(x)$ variation.

The boundary conditions (\ref{Theta boundary})--(\ref{Velocity boundary}) can be rewritten in terms of the dimensionless variable $z=x/\xi$ as
\begin{gather}
     \frac{d\theta(\pm0,\omega)}{dz}=\pm\frac{\alpha}{2}\sin2\theta(0,\omega)\left[1-\cos\delta\chi(\omega)\right],{\label{DimensionlessBCTheta}}\\
    \frac{d\chi(\pm0,\omega)}{dz}=\alpha\sin\delta\chi(\omega){\label{DimensionlessBCVelocity}},
\end{gather}
where we have defined the dimensionless conductance parameter
\begin{equation} \label{AlphaDef}
    \alpha = \xi g_N / \sigma.
\end{equation}
This parameter can be rewritten as \cite{Belzig1999}
\begin{equation} \label{alpha_t}
    \alpha=2\xi t / l,
\end{equation}
where the average barrier transparency $t$ is small in the tunneling limit.

Due to finite value of $\alpha$, the proximity effect between the two sides of the Josephson junction leads to suppression of $|\Delta(x)|$ in the vicinity of the interface (at nonzero phase difference).
We standardly define the tunneling limit as the regime in which the proximity effect [i.e., suppression of $|\Delta(x)|$] is weak. This condition implies that $\alpha$ must be small. The exact condition for the smallness of $\alpha$ will be discussed below in Sec.~\ref{sec:applicability}.

One more point regarding various interface parameters should be commented here.
The KL boundary conditions (\ref{Theta boundary}) and (\ref{Velocity boundary}) are valid in the limit of small transparencies of interface conducting channels, which may be formulated as $t\ll 1$. They can be obtained in the first order with respect to $t$ from the more general Nazarov boundary conditions \cite{Nazarov1999}.
We plan to do the perturbation theory with respect to $\alpha$ (staying in the regime of validity of the KL boundary conditions) but we do not take into account higher-order terms with respect to $t$ from the Nazarov boundary conditions.
This is legitimate since $\alpha \gg t$ [see Eq.\ (\ref{alpha_t})] due to the diffusive limit condition $\xi\gg l$.

For example, the next-order term from the Nazarov boundary conditions would lead to contributions of the order of $\alpha t$ in the right-hand sides of Eqs.\ (\ref{Theta boundary}) and (\ref{Velocity boundary}) (and in the solutions).
At the same time, the proximity effect treated within the KL boundary conditions leads to corrections of the order of $\alpha^2$. Since $\alpha^2 \gg \alpha t$, the main effect is captured by the self-consistent theory based on the KL boundary conditions.

\section{Perturbation theory with respect to the interface conductance: Josephson current}
\label{sec:three}

\subsection{Arbitrary temperatures}

The starting point of our perturbation theory is the solution of the Usadel equations with the KL boundary conditions at $\alpha=0$.
This trivial solution can be written as
\begin{gather}
    |\Delta(z)|=\Delta_{0}\label{Delta_zero},\\
    \theta(z,\omega)=\theta_{0}\equiv\arctan ( \Delta_{0} / \omega) {\label{Theta_0}},\\
    \chi(z,\omega)=\varphi(z)= (\delta\varphi /2) \sgn z. \label{Phi_0}
\end{gather}
We consider the order-parameter phase jump at the interface,
\begin{equation}
\delta\varphi=\varphi(+0)-\varphi(-0),
\end{equation}
as the parameter that defines the current-carrying state of the Josephson junction.
This parameter enters the full self-consistent set of equations and determines, in particular, the strength of the proximity effect between the superconducting banks and the current at any point of the junction.

Expanding $\theta$, $\chi$, $\Delta$, and $\varphi$ in powers of $\alpha$, we get
\begin{align}
    \theta(z,\omega)&=\theta_{0}+\alpha\theta_{1}(z,\omega)+\alpha^2\theta_{2}(z,\omega),{\label{Theta_pert}}\\
    |\Delta(z)|&=\Delta_{0}+\alpha\Delta_{1}(z)+\alpha^2\Delta_{2}(z),{\label{Delta_pert}}\\
    \chi(z,\omega)&= (\delta\varphi /2) \sgn{z}+\alpha\chi_{1}(z,\omega)+\alpha^{2}\chi_{2}(z,\omega),{\label{Chi_pert}}\\
    \varphi(z)&= (\delta\varphi/ 2) \sgn{z}+\alpha\varphi_{1}(z)+\alpha^{2}\varphi_{2}(z),{\label{Phi_pert}}\\
    \delta\chi(\omega)&=\delta\varphi+\alpha\delta\chi_1(\omega)+\alpha^2\delta\chi_2(\omega).
\end{align}

The phases $\chi(z,\omega)$ and $\varphi(z)$ in the case of nonzero current $J\neq0$ grow linearly in the bulk of the superconductors. Corrections $\chi_{1(2)}$ and $\varphi_{1(2)}$ therefore become large, which may seem to create a problem for our perturbation theory. However, this problem is purely formal because the quantities that actually enter our perturbation theory are not $\chi$ and $\varphi$ themselves but their derivatives $d\chi/dz$ and $d\varphi/dz$ as well as their difference $\chi-\varphi$; all those quantities are finite in the bulk.

Our goal is to find the answer for $J$ up to the $\alpha^2$ order. The current given by Eq.\ (\ref{Current relation}) contains $d\chi/dz\sim\alpha$; so, in order to obtain the answer up to $\alpha^2$, it is sufficient to find $\theta_1$, $\Delta_1$, and $\chi_2$.

We start with calculating $\theta_1$ and $\Delta_1$. In the first order of the perturbation theory, equations for $\theta_1$ and $\Delta_1$ separate from equations for $\chi_1$ and $\varphi_1$, the pair-breaking term $(d\chi/dx)^2$ in Eq.\ (\ref{Diffusion}) should be dropped out, and $\cos(\chi-\varphi)$ should be substituted by $1$.

The Usadel equation (\ref{Diffusion}) and the boundary condition (\ref{Theta boundary}) up to the first power in $\alpha$ have the form
\begin{gather}
        \frac{d^{2}\theta_{1}}{dz^{2}}+\frac{\Delta_{1}(z)}{\Delta_{0}}\cos\theta_{0}-\frac{\theta_{1}}{\sin\theta_{0}}=0,{\label{Diffusion linearized}}\\
            \frac{d}{dz}\theta_{1}(\pm0)=\pm\frac{1}{2} \left(1-\cos\delta\varphi\right) \sin2\theta_{0}.\label{Theta boundary linearized}
\end{gather}
We can solve this linear system with the help of the Fourier transformation (with respect to $z$). In the Fourier space we find
\begin{equation}\label{Theta-Delta connection}
    \theta_{1}(k)=\frac{\sin\theta_{0}\cos\theta_{0}}{k^{2}\sin\theta_{0}+1}\left[ \frac{\Delta_{1}(k)}{\Delta_{0}}-2\left(1-\cos\delta\varphi\right)\sin\theta_{0}\right].
\end{equation}

In the first order of the perturbation theory, the real part of the self-consistency equation (\ref{Self-consistency}) yields
\begin{equation}{\label{Self-consistency linearized}}
\Delta_{1}(k)=\pi\lambda T\sum_{|\omega|<\omega_{D}}\theta_{1}(k,\omega)\cos\theta_{0}(\omega).
\end{equation}

The answer for $\Delta_1$ can be  written in terms of $\Delta_0$ without any explicit information on $\omega_D$ and $\lambda$.
The bulk self-consistency equation can be written as
\begin{equation}\label{Weak-link substitution}
    \frac{1}{\lambda}=\frac{\pi T}{\Delta_{0}}\sum_{|\omega|<\omega_{D}}\sin\theta_{0}.
\end{equation}
Substituting this expression for $\lambda$ into Eq.\ (\ref{Self-consistency linearized}), we can rewrite the latter equation in the form
\begin{equation}\label{NoLogSelfConsist}
    \pi T\sum\limits_{|\omega|<\omega_D}\left(\frac{\Delta_1}{\Delta_0}\sin\theta_0-\theta_1\cos\theta_0\right)=0.
\end{equation}
Plugging expression (\ref{Theta-Delta connection}) for $\theta_1$ into this equation, we see that the resulting sum is convergent. We can therefore extend the Matsubara summation to infinite limits (formally putting $\omega_D=\infty$). For more details, see Appendix \ref{Appendix B}.

We introduce the following notation for a class of sums arising as a result of this procedure:
\begin{equation}{\label{Sums}}
    L_{n}(k,T)\equiv\frac{2\pi T}{\Delta_{0}}\sum_{\omega>0}\frac{\sin^{n}\theta_{0}}{k^{2}\sin\theta_{0}+1}.
\end{equation}
The result for $\Delta_1 (k)$ can then be written in the form
\begin{equation}{\label{Delta_1}}
\frac{\Delta_{1}(k)}{\Delta_{0}}=-2(1-\cos\delta\varphi)\frac{L_{2}(k,T)-L_{4}(k,T)}{k^{2}L_{2}(k,T)+L_{3}(k,T)}.
\end{equation}

\begin{figure}[t]
\centerline{\includegraphics[width=\columnwidth]{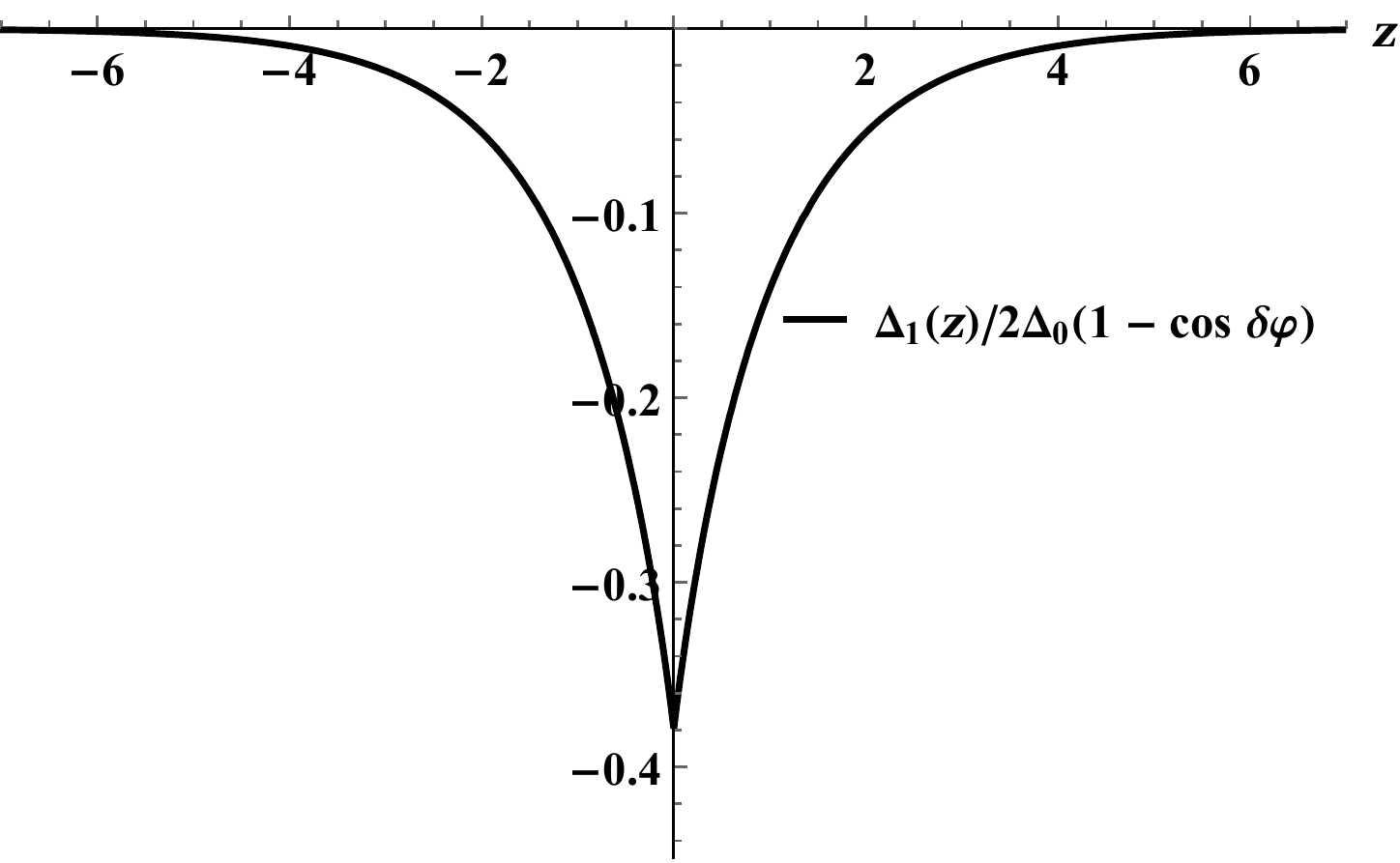}}
\caption{$\Delta_1(z)/\Delta_0$ plot at $T=0$ without factor $2(1-\cos\delta\varphi)$. The correction to the order parameter is negative and the order parameter is most strongly suppressed in the vicinity of the interface.}
\label{FigureDelta1}
\end{figure}

Figure~\ref{FigureDelta1} illustrates the correction to the order parameter $\Delta_1(z)/\Delta_0$ in the coordinate space at $T=0$. Since $\Delta_1$ is proportional to $2(1-\cos\delta\varphi)$, the plot is shown without this factor. As one can see, the result of calculations is in line with the expectations shown schematically in Fig.~\ref{FigureSIS}.

The next step is to find $\chi_1$ and $\varphi_1$. This can be done by using the linearized form of the continuity equation (\ref{Continuity}), the imaginary part of the self-consistency equation (\ref{Self-consistency}), and the boundary condition (\ref{Velocity boundary}) for the velocity of the Cooper pairs,
\begin{gather}
\frac{d^{2}\chi_{1}}{dz^{2}} \sin^{2}\theta_{0}=\left(\chi_{1}-\varphi_{1}\right)\sin\theta_{0}{\label{Continuity linearized}},\\
    \sum_{|\omega|<\omega_{D}}\left(\chi_{1}-\varphi_{1}\right)\sin\theta_{0}=0{\label{ImSelf-consistency linearized}},\\
    \frac{d}{dz}\chi_{1}(\pm0)=\sin\delta\varphi.{\label{Velocity boundary linearized}}
\end{gather}

The solution of this system is trivial,
\begin{equation}{\label{Phase equality}}
    \chi_{1}(z,\omega)=\varphi_{1}(z)=z\sin\delta\varphi.
\end{equation}
This formula tells us that in the main order with respect to the interface conductance, the Josephson relation have the standard form $J\propto \sin\delta\varphi$. Moreover, $\chi_1(z,\omega)=\varphi_1(z)$ are continuous functions at $z=0$, unlike $\chi_0(z,\omega)=\varphi_0(z)=(\delta\varphi/2)\sgn{z}$. Therefore, $\delta\chi_1(\omega)=0$.

Expanding Eq.\ (\ref{DimensionlessBCVelocity}) up to $\alpha^2$, we obtain
\begin{equation}{\label{Second order velocity boundary}}
    \frac{d}{dz}\chi_{2}(\pm0)=0.
\end{equation}
This boundary condition implies that in order to calculate $J$ (which can be done at $z=0$), we do not actually need to calculate $\chi_2(z)$.
To find the current up to the $\alpha^2$ order, we thus only need $\Delta_1$ and $\theta_1$. For more details, see Appendix \ref{Appendix B}.

The answer for the current has the form
\begin{gather} \label{Current answer}
    J=J_{0}\sin\delta\varphi\left[ 1-4\alpha\left(1-\cos\delta\varphi\right)V(T)\right],\\
    J_{0}\equiv\frac{\pi G_N \Delta_{0}}{2e}\tanh\left(\frac{\Delta_{0}}{2T}\right),\\
    V(T)=\coth\left(\frac{\Delta_{0}}{2T}\right)\int\limits_{-\infty}^{\infty}\frac{dk}{\pi^2}\left[\frac{\left(L_{2}-L_{4}\right)^{2}}{k^{2}L_{2}+L_{3}}+(L_{3}-L_{5}) \right]\label{V(T)},
\end{gather}
where $V(T)$ is a positive number, which depends on temperature $T$. While at arbitrary temperature, $V(T)$ can be found only numerically, we can find explicit results in the limiting cases of $T\to  0$ and $T\to  T_{c}$.

The answer for the current given by Eq.\ (\ref{Current answer}) contains not only the standard part of the Josephson relation $J_0 \sin\delta\varphi$ but also the second harmonic ($\sin 2\delta\varphi$ with positive coefficient) and a negative correction to the first harmonic. Below we present results in the limiting cases of $T=0$ and $T\to  T_c$.

\subsection{Limiting cases}
\label{sec:limits}

In the limiting cases, we find
\begin{equation} \label{V(T)Answer}
      V(T)=
      \left\{
        \begin{aligned}
        &\frac{\pi}{2}\left[\frac{1}{56\zeta(3)\left(1-T/T_{c}\right)}\right]^{1/4}, && T\to  T_{c}, \\
        &0.272, && T\to 0.
        \end{aligned}
        \right.
\end{equation}

Thus the answer for the current at $(T_c-T)\ll T_c$ is
\begin{gather}\label{CurrentTcAnswer}
    J=\frac{\pi G_N \Delta_{0}^{2}(T)}{4e T_{c}}\left\{ \left[1- \sqrt{2} \gamma(T) \right] \sin\delta\varphi+\frac{\gamma(T)}{\sqrt{2}}\sin2\delta\varphi \right\},\\
    \gamma(T) = G_N / G_{D}(T), \label{gamma}
\end{gather}
where $G_D(T) = \sigma S / \xi_\mathrm{GL}(T)$ is the diffusive conductance of the superconductor on the length $\xi_\mathrm{GL}(T)$.

In the $T\to 0$ limit, the characteristic length scale becomes $\xi(0)$. At the same time, since $\xi(0) \sim \xi_\mathrm{GL}(0)$, we can write the answer with the help of the same definition for $G_D(T)$ as
\begin{multline}\label{CurrentT0Answer}
    J=\frac{\pi G_N \Delta_{0}(0)}{2e}
    \\
    \times \left\{ \left[ 1 - 0.93\gamma(0) \right] \sin\delta\varphi+0.46\gamma(0) \sin2\delta\varphi\right\}.
\end{multline}

\subsection{Comparison with previous results}
\label{sec:previous results}
Equations (\ref{Current answer})--(\ref{CurrentT0Answer}) are the main results of this paper.
In the limit $T\to  T_c$, Eq.\ (\ref{Current answer}) reduces to Eq.\ (\ref{CurrentTcAnswer}) and reproduces the result by Kupriyanov \cite{Kupriyanov1992}. The answer (\ref{Current answer}) for arbitrary $T$ and its $T\to  0$ limit, Eq.\ (\ref{CurrentT0Answer}),  are
new results.

The Josephson relation with the second harmonic in the SIS tunnel junction at arbitrary $T$ was previously derived by Golubov and Kupriyanov \cite{GK2005}. In their paper, the Usadel equations (\ref{Diffusion})--(\ref{Self-consistency}), (\ref{Theta boundary}), and (\ref{Velocity boundary}) were solved in the coordinate space. The authors employed a conjecture for the form of the solution to the full self-consistent problem
\footnote{In Ref.\ \cite{GK2005}, the perturbation theory was developed in the coordinate space. Two issues indicate that the presented form of solution is not rigorous (we call it ``conjectured''). (i)~In Eq.\ (31) of  Ref.\ \cite{GK2005}, the order parameter and the quantity parametrizing the Green functions are expanded in the system of decaying exponents. The system does not form a full basis in the functional space, which means that actually only a certain class of functions is considered.
(ii)~Equation (34) in Ref.\ \cite{GK2005} is obtained from Eqs.\ (32) and (33) according to the procedure described below Eq.\ (33). This procedure leads to equality between two sums running over \emph{different} quantities ($\omega$ and $\Omega$). In order to obtain Eq.\ (34), one should equate term-by-term the elements of these \emph{different} sums. This assumption also implies a certain conjecture about the form of solution.}.
On the contrary, our perturbation theory allows systematic rigorous calculation of the solution. The results of Ref.\ \cite{GK2005} for the Josephson current turn out to be parametrically correct but with wrong numerical coefficients in front of the $\gamma$ corrections. Our theory provides exact values of the coefficients.

The result of Ref.\ \cite{GK2005} for the current can be written in the form of Eq.\ (\ref{Current answer}) but with a different coefficient $V_\text{GK}(T)$ instead of $V(T)$. We can therefore characterize the magnitude of difference between our final results by comparing the two quantities.
In the $T=0$ limit, expressions from  Ref.\ \cite{GK2005} imply $V_\text{GK}(0)=B(3/2,3/4)/\pi\approx 0.305$ [where $B(x,y)$ is the Euler beta function], instead of our value $V(0) \approx 0.272$, see Eq.\ (\ref{V(T)Answer}). In the $T\to T_c$ limit, $V_\text{GK}(T)=16V(T)/\pi^2$.
So, the difference amounts to a factor, which can exceed $1.6$.

\section{Perturbation theory for the phases}
\label{sec:four}

As we have shown in Eq.\ (\ref{Phase equality}) in the first order of the perturbation theory, the phases $\chi_1(z,\omega)$ and $\varphi_1(z)$ are the same at all frequencies. In this section, we show that the second-order perturbation theory yields $\chi_2(x,\omega)\neq\varphi_2(x)$.

We start the second-order perturbation theory by expanding Eqs.\ (\ref{Continuity}), (\ref{Self-consistency}), and (\ref{DimensionlessBCVelocity}) up to $\alpha^2$. Thus, we obtain
\begin{gather}
    \sin\theta_{0}\frac{d^{2}\chi_{2}}{dz^{2}}+2\frac{d\theta_{1}}{dz}\cos\theta_{0}\sin\delta\varphi=\chi_{2}-\varphi_{2},\label{ContinuityDoubleLinearized}\\
    \sum_{|\omega|<\omega_{D}} \left[ \chi_2(z,\omega)-\varphi_{2}(z)\right] \sin\theta_{0} (\omega)=0,\label{SelfConsistencyDoubleLinearized}\\
    \frac{d}{dz}\chi_{2}(\pm0,\omega)=0.\label{VelocityBoundary2}
\end{gather}
Due to linearity of the system, we solve it with the help of the Fourier transformation. We must take into account that $\chi_2(z,\omega)$ can be discontinuous at $z=0$ with a (yet unknown) phase jump $\delta\chi_2(\omega)$. Moreover, in the bulk $d\chi_2/dz$ can be finite, so it is convenient to write equations for new variable
\begin{equation}
    \phi_2(z,\omega)\equiv\chi_2(z,\omega)-\varphi_2(z),
\end{equation}
which has zero derivative in the bulk, where $\chi_2=\varphi_2$.
In the Fourier space, we obtain
\begin{gather}
    \phi_{2}(k,\omega)= ik\frac{(ik\varphi_{2}-\delta\chi_{2})\sin\theta_{0}+2\theta_{1}\cos\theta_{0}\sin\delta\varphi}{1+k^{2}\sin\theta_{0}} ,\label{ContinuityDoubleFourier}\\
    \sum_{|\omega|<\omega_{D}} \phi_2(k,\omega)\sin\theta_{0}=0,\label{SelfConsistencyDoubleLFourier}\\
    \lim_{z\to 0}\biggl( \int\limits_{-\infty}^{\infty}\frac{dk}{2\pi}ik\phi_{2}e^{ikz}-\delta\chi_{2}\delta(z) \biggr)=0.\label{PhasesFourierBoundary}
\end{gather}
The system of Eqs.\ (\ref{ContinuityDoubleFourier})--(\ref{PhasesFourierBoundary}) determines $\phi_2(z,\omega)$ and $\delta\chi_2(\omega)$. In order to find these functions, we employ an algorithm similar to the one used in the case of $\theta$ and $\Delta$. First, we substitute Eq.\ (\ref{ContinuityDoubleFourier}) into Eq.\ (\ref{SelfConsistencyDoubleLFourier}) and then find $ik\varphi_2(k)$ and  $\phi_2(k)$. We still do not know $\delta\chi_2(\omega)$ but we can find it from Eq.\ (\ref{PhasesFourierBoundary}). This procedure gives us the following equation for $\delta\chi_2(\omega)$ (for more details, see Appendix \ref{Appendix C}):
\begin{gather}\label{PhaseShiftEquation}
    \frac{\delta\chi_{2}}{2\sqrt{\sin\theta_{0}}}=\frac{d\chi_{2}^{(0)}(z=0)}{dz}-V_0+\int\limits_{-\infty}^{\infty}\frac{dk}{2\pi}\frac{\Phi(k)/L_2(k)}{1+k^{2}\sin\theta_{0}},\\
    V_0=4V(1-\cos\delta\varphi)\sin\delta\varphi,
\end{gather}
where $\chi_{2}^{(0)}(z,\omega)$ and  $\varphi_{2}^{(0)}(z)$ are the auxiliary functions that solve the system of Eqs.\ (\ref{ContinuityDoubleLinearized}) and (\ref{SelfConsistencyDoubleLinearized}) with the (auxiliary) conditions that $\chi_{2}^{(0)}$ is a continuous function of $z$ vanishing in the bulk. The constant $V_0$ is a new constant, proportional to $V$.
Since $V_0=-d\varphi_2(z=\infty)/dz$, then $V_0$ determines the correction to the velocity of Cooper pairs in the bulk according to $d\chi/dz =\alpha\sin\delta\varphi- \alpha ^2V_0$ (for more details, see Appendix \ref{Appendix C}). Here we also define the phase functional $\Phi(k)$ according to
\begin{equation}\label{PhaseFunctional}
    \Phi[\delta\chi_{2}](k)\equiv\frac{2\pi T}{\Delta_{0}}\sum_{\omega>0}\frac{\delta\chi_{2}\sin^{2}\theta_{0}}{1+k^{2}\sin\theta_{0}}.
\end{equation}
The solution of Eq.\ (\ref{PhaseShiftEquation}) gives us $\delta\chi_2(\omega)$ and $\Phi(k)$, and with the use of Eqs.\ (\ref{ContinuityDoubleFourier}) and (\ref{SelfConsistencyDoubleLFourier}), we can find $\chi_2(z,\omega)$ and $\varphi_2(z)$.

Even without explicit implementation of this algorithm, we can make sure that $\chi_2\neq \varphi_2$.
Indeed, if we assume that $\chi_2=\varphi_2$, then Eq.\ (\ref{PhaseShiftEquation}) immediately simplifies to the form
\begin{equation}
    \frac{d\chi_{2}^{(0)}(0,\omega)}{dz}=V_0,
\end{equation}
which cannot be satisfied since both $\chi_{2}^{(0)}(z,\omega)$ and its derivative at $z=0$ have nontrivial dependence on $\omega$ (as witnessed, for example, by numerical calculations).
This proves that $\chi_2\neq\varphi_2$. Moreover, this result is a consequence of Eq.\ (\ref{ContinuityDoubleLinearized}), which contains  $d\theta_1/dz$ that plays the role of the nonzero source in this equation.

\begin{figure}[t]
\centerline{\includegraphics[width=\columnwidth]{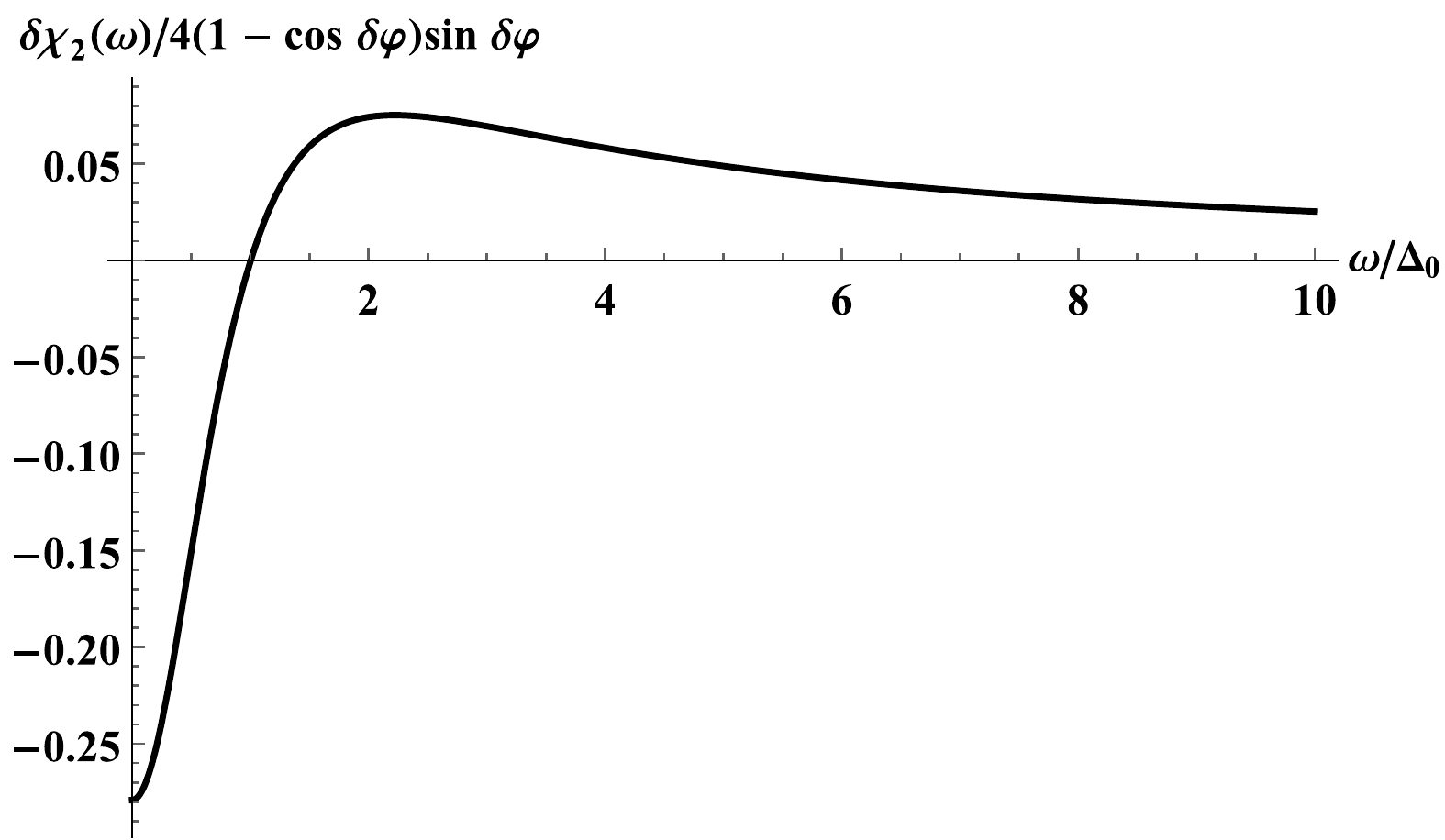}}
\caption{$\delta\chi_2(\omega)$ plot at $T=0$ without factor $4(1-\cos\delta\varphi)\sin\delta\varphi$.
Interestingly, the curve crosses the abscissa very close to the $\omega=\Delta_0$ point. While this may be a hint to an exact property, we do not have a proof for that.}
\label{FigureDeltaChi2}
\end{figure}

\begin{figure}[t]
\centerline{\includegraphics[width=\columnwidth]{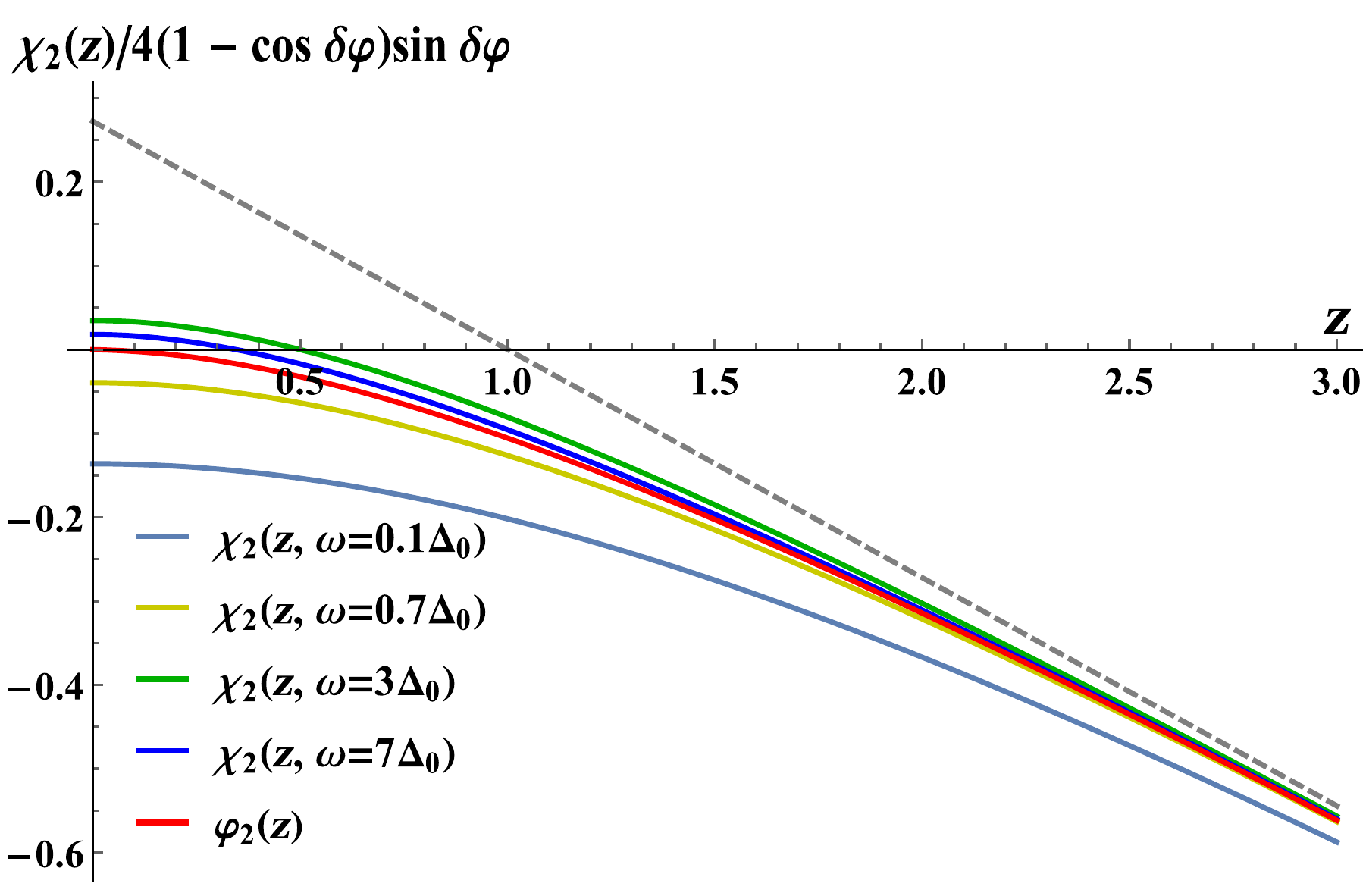}}
\caption{$\chi_2(z)$ plot at different $\omega$ without factor $4(1-\cos\delta\varphi)\sin\delta\varphi$. Since $\chi_2$ is an odd function, nonzero values $\chi_2(z=0)\neq 0$ signify that this function is discontinuous at $z=0$. At $z\to \infty$, all the curves become linear. Interestingly, this linear dependence of the form $a\sgn z+bz$ crosses the $z$ axis very close to $z=1$. While this may be a hint to an exact property, we do not have a proof for that. The figure demonstrates that $\delta\chi_2$ changes nonmonotonically as the function of $\omega$: the curve for $\omega=7\Delta_0$ is lower than for $\omega=3\Delta_0$, but higher than for $\omega=0.7\Delta_0$ (this correlates with the result of Fig.~\ref{FigureDeltaChi2}). In the $\omega\to \infty$ limit, the $\chi_2(z,\omega)$ curves converge to $\varphi_2(z)$.}
\label{FigurePhasesRealSpace}
\end{figure}

Equation (\ref{PhaseShiftEquation}) can be solved numerically, and we present the results of this procedure in the case of $T=0$ in Figs.~\ref{FigureDeltaChi2} and~\ref{FigurePhasesRealSpace}. Both the figures confirm that $\chi\neq\varphi$. From Fig.~\ref{FigureDeltaChi2}, we see that $\delta\chi_2(\omega)$ is an alternating function, which could be inferred from Eq.\ (\ref{SelfConsistencyDoubleLinearized}) at $z=\pm 0$. Indeed, due to the continuity of corrections $\varphi_1(z)$ and $\varphi_2(z)$, we obtain
\begin{equation}
    \sum_{\omega}\delta\chi_2\sin\theta_0=0.
\end{equation}
The sum can turn to zero only if $\delta\chi_2(\omega)$ changes its sign.

Figure \ref{FigurePhasesRealSpace} demonstrates how the phases $\chi_2(z,\omega)$ and $\varphi_2(z)$ depend on $z$ in a nonlinear manner (such nonlinear dependence was discussed in Ref.\ \cite{Ivanov1981Eng} in the case of SNS junction). $\chi_2(z,\omega)$ and $\varphi_2(z)$ become equal in the bulk and vary linearly with the slope $-V_0(T)$,
\begin{equation}\label{LinearLaw}
	\chi_{2}(z\to\pm\infty,\omega)=\varphi_{2}(z\to\pm\infty)=a\sgn z-V_{0}z,
\end{equation}
where $a$ is a constant, which can be obtained after solving the integral equation (\ref{PhaseShiftEquation}) (for more details, see Appendix \ref{Appendix C}).

Physically, the overall nonlinear spatial dependence of the phases corresponds to increased velocity of the superconducting condensate in the vicinity of the interface.
This compensates for the interface suppression of the order parameter (see Fig.~\ref{FigureDelta1}) and, hence, of the condensate density (due to the proximity effect between the superconducting banks with different phases) in order to provide position-independent Josephson current throughout the system.

More details on the second-order perturbation theory for the phases are presented in Appendix \ref{Appendix C}.

\section{Discussion}
\label{sec:Discussion}

\subsection{Applicability conditions of the perturbation theory}
\label{sec:applicability}

The condition of weak proximity effect, which we assumed when developing our perturbation theory, can be formulated according to Eq.\ (\ref{Delta_pert}) as
\begin{equation}
 \alpha |\Delta_1(z=0)|/\Delta_0 \ll 1.
\end{equation}
In the Fourier representation, the result for $\Delta_1$ is given by Eq.\ (\ref{Delta_1}). At $T$ not too close to $T_c$, this yields $|\Delta_1(z=0)| \sim \Delta_0$, so that the condition becomes $\alpha \ll 1$.
At the same time, at $T\to T_c$, Eq.\ (\ref{Delta1(k)}) demonstrates that $|\Delta_1(z=0)| / \Delta_0 \sim \sqrt{T_c / \Delta_0(T)} \sim (1-T/T_c)^{-1/4}$.

Summarizing, at all temperatures, the condition of smallness of $\alpha$ can be written as
\begin{equation} \label{alphasmall}
\alpha \ll (1-T/T_c)^{1/4}.
\end{equation}
Note that the $\alpha$ parameter itself depends on $T$ in the vicinity of $T_c$ as $\alpha \propto (1-T/T_c)^{-1/4}$.

Alternatively, condition (\ref{alphasmall}) can be written as $\gamma (T)\ll 1$, where $\gamma(T)$ is defined by Eq.\ (\ref{gamma}). The limiting results for the Josephson current, Eqs.\ (\ref{CurrentTcAnswer}) and (\ref{CurrentT0Answer}), confirm that this is indeed the condition of smallness of the corrections to the Josephson relation.

\subsection{Role of self-consistency}

While self-consistency for the order parameter is inherent in our calculations, it may be instructive to discuss its role, considering what changes if the self-consistency is neglected and we simply put $\Delta = \Delta_0$. Below, we discuss how this would change the results for the Josephson current $J$.

Neglecting self-consistency implies putting $\Delta_1(k)=0$ in Eq.\ (\ref{Theta-Delta connection}). Following step-by-step the algorithm described in Appendix \ref{Appendix B}, one would then obtain
\begin{equation} \label{Vwithout_sc}
V(T)=\coth\left(\frac{\Delta_{0}}{2T}\right)\int\limits_{-\infty}^{\infty}\frac{dk}{\pi^2}\left(L_{3}-L_{5}\right).
\end{equation}
Neglecting self-consistency thus leads to dropping out the first term under the integral in Eq.\ (\ref{V(T)}).

In the limit of low temperatures, $T\to 0$, neglecting self-consistency makes the result for the numerical coefficient $V$, Eq.\ (\ref{Vwithout_sc}), valid only by the order of magnitude.
Indeed, for the frequencies $\omega\sim\Delta_0$, we have $\sin\theta_0\sim 1$, which means that the $L_n$ sums defined in Eq.\ (\ref{Sums}) are all of the same order and vary on the scale of $k \sim 1$. Therefore, both the terms under the integral in Eq.\ (\ref{V(T)}) are of the same order.

In the case of approaching the critical temperature, $T\to  T_c$, self-consistency begins to play a major role. Indeed, in this limit, we have $\sin\theta_0\ll 1$, hence the $L_n$ sums are of the order of $(\Delta_0/T_c)^{n-1}$ and vary on the scale of $k\sim \sqrt{T_c/\Delta_0} \gg 1$. Substituting this into the integral in Eq.\ (\ref{V(T)}), one finds that the first term (which is due to self-consistency) gives a contribution of the order of $\sqrt{T_c/\Delta_0}$, while the $(L_3-L_5)$ term gives a contribution of the order of $\sqrt{\Delta_0/T_c}\ll\sqrt{T_c/\Delta_0}$. The major role of self-consistency in this case is expectable since in the $T\to  T_c $ limit, the Usadel equations reduce to the GL equations \cite{Svidzinsky1982}, so all information about spatial variations of superconducting characteristics inside the superconducting banks must be encoded in the $\Delta(x)$ function. Neglecting this spatial dependence would mean neglecting the corrections due to finite interface conductance, which is the main effect considered in this paper.

We thus conclude that taking into account self-consistency in our problem is necessary in order to obtain quantitatively and qualitatively correct results.

\subsection{Different definitions of the phase jump}
\label{sec:phase jump definiton}

Physically, the current-carrying state of the SIS junction can be defined in several ways. In this paper, we parametrize it by $\delta\varphi$, the order-parameter phase jump at the interface. At the same time, it can be more experimentally relevant \cite{Ivanov1981Eng} to define the phase jump 
not at the interface but in the bulk of the superconductor after subtracting the linearly-growing contribution,
\begin{equation}
	\delta\varphi_{\text{nonlin}}\equiv2\left[ \varphi(z\to\infty)-z\frac{d\varphi(z\to\infty)}{dz}\right],
\end{equation}
where the factor of $2$ takes into account that $\varphi(z)$ is an odd function. Equation (\ref{LinearLaw}) provides the connection between $\delta\varphi$ and $\delta\varphi_{\text{nonlin}}$,
\begin{equation} \label{deltaphi-deltaphinonlin}
\delta\varphi_{\text{nonlin}}=\delta\varphi+2\alpha^2 a.
\end{equation}
While we study $J(\delta\varphi)$, the current-phase relation could alternatively be defined as $J(\delta\varphi_{\text{nonlin}})$. Nevertheless, within our accuracy, this difference does not change any of the above results.
This is because in the Josephson relation Eq.\ (\ref{Current answer}), the difference defined by Eq.\ (\ref{deltaphi-deltaphinonlin}) would manifest itself only in the third order with respect to $\alpha$, which we do not consider (since $J_0$ is itself of the first order).

\subsection{Outlook}
\label{sec:outlook}

We have shown that the phases $\varphi$ and $\chi$ differ in the second order of the perturbation theory, but this difference does not immediately show up in the answer for the current, Eq.\ (\Ref{Current answer}), due to the boundary condition for $\chi_2$, Eq.\ (\ref{Second order velocity boundary}). This is because we calculate the current at the ``symmetric'' point of the SIS system, the interface (the answer does not depend on the point at which it is calculated). At the same time, the $\varphi$--$\chi$ difference would directly influence the calculation of current at any finite distance from the interface.
On the other hand, we expect that the $\varphi$--$\chi$ difference would immediately show up in the answer for the current at any point (including the interface) in the third and higher orders of the perturbation theory.

At the same time, the Josephson current is an integral quantity (the sum of the spectral components of the current), and one can therefore expect the $\varphi$--$\chi$  difference to manifest itself more clearly (both from theoretical and experimental point of view) in spectrally-resolved quantities. The most obvious quantity of this sort is the local density of states in the vicinity of the interface, which requires reformulating the theory in the real-energy technique. The behavior of the system in the alternating electric field, so-called Shapiro steps, should also be sensitive to spectrally-resolved characteristics of the system.

In addition, an interesting direction is to study systems such as SNS and SFS junctions (where N is a normal metal and F is a ferromagnet).
While in the case of tunnel SIS junctions the second Josephson harmonic (which we have calculated) is always small compared to the first one, the SFS case looks especially intriguing since the first harmonic can be suppressed in this case due to specific nature of proximity-induces superconductivity inside the F interlayer \cite{Golubov2004review}.

We leave the above questions for future studies.

\section{Conclusions}
\label{sec:Conclusions}

We have considered the Josephson effect in a planar diffusive SIS-type junction at arbitrary temperature and constructed fully self-consistent perturbation theory with respect to the dimensionless conductance parameter $\alpha \ll 1$, which is the ratio of the interface conductance to the conductance of the superconducting material on the coherence length. We have presented analytical analysis of two orders of the perturbation theory.

The first order of the perturbation theory provides correction $\Delta_1$ to the absolute value of the order parameter, see Eq.\ (\ref{Delta_1}) and Fig.~\ref{FigureDelta1}. In the coordinate space, $\Delta(z)$ is suppressed in the vicinity of the interface. Knowledge of $\Delta_1$ makes it possible to find $\theta_1$. In its turn, $\theta_1$ provides the answer for the Josephson current up to the $\alpha^2$ order, which contains not only the standard part $J(\delta\varphi) \propto\sin\delta\varphi$ but also a (negative) correction to the first harmonic and the second harmonic $\sin 2 \delta\varphi$ (with a positive amplitude). We further analyze the general answer given by Eq.\ (\ref{Current answer}), in two limiting cases, see Sec.~\ref{sec:limits}. In the $T\to  T_c$ limit, we reproduce the result by Kupriyanov \cite{Kupriyanov1992}, while our results in the $T\to 0$ limit (as well as in the case of arbitrary temperature) have not been reported before. Although the same problem at arbitrary temperature has been considered before in Ref.\ \cite{GK2005}, the corrections to the Josephson relation obtained there were only parametrically correct due to a conjectured form of solution. Our theory provides rigorous solution, which results in exact numerical coefficients.

Our perturbation theory also provides solutions for the superconducting phases of the anomalous Green functions and of the order parameter, $\chi$ and $\varphi$, respectively.
In the zeroth order, the phases are equal constants corresponding to the standard main-order solution for the Josephson effect in tunnel junctions.
In the first order, the phases are still equal but acquire the linear part, which describes finite velocity of the superconducting condensate at each point of the superconductors.
Finally, in the second order, we find that $\chi\neq\varphi$. We present the plot of $\chi_2(z,\omega)$ at different Matsubara frequencies and of $\varphi_2(z)$ at $T=0$ in Fig.~\ref{FigurePhasesRealSpace}. We also illustrate the frequency dependence of the phase jumps $\delta\chi_2(\omega)$ at $T=0$ in Fig.~\ref{FigureDeltaChi2} (note that the phase jumps $\delta\varphi_2$ are absent by definition).

The overall spatial dependence of the phases is nonlinear, corresponding to increased velocity of the superconducting condensate in the vicinity of the interface.
This compensates for the interface suppression of the order parameter and, hence, of the condensate density (due to the proximity effect between the superconducting banks) in order to provide position-independent Josephson current throughout the system, see Fig.~\ref{FigureSIS}.

\acknowledgments
We thank A.~A.\ Golubov, M.~V.\ Feigel'man, P.~M.\ Ostrovsky, and especially M.~Yu.\ Kupriyanov for useful discussions.
The work was supported by
the Russian Science Foundation (Grant No.\ 21-42-04410)
and
the Foundation for the Advancement of Theoretical Physics and Mathematics ``BASIS''.


\appendix

\section{Equations and parametrization}\label{Appendix A}

The Usadel equations are written for the isotropic (in the momentum space) parts of the quasiclassical Green functions, $G$ and $F$ \cite{Usadel1970},
\begin{gather}
    \frac{D}{2}\left(G\nabla^{2}F-F\nabla^{2}G\right)+G\Delta-\omega F=0,
    \label{Usadel} \\
    \frac{D}{2}\left(G\nabla^{2}\tilde{F}-\tilde{F}\nabla^{2}G\right)+G\Delta^{*}-\omega\tilde{F}=0,
    \label{CCUsadel} \\
    D \nabla \left(G\nabla G +F \nabla \tilde{F} \right) - \tilde{F} \Delta + F \Delta^* =0,
    \label{Usadel3}\\
    G^{2}+F\tilde{F}=1. \label{NC}
\end{gather}

The third equation, Eq.\ (\ref{Usadel3}), is actually a consequence of the first two equations and the normalization condition (\ref{NC}). In order to see that, one can multiply Eqs.\ (\ref{Usadel}) and (\ref{CCUsadel}) by $\tilde F$ and $F$, respectively, and consider the difference between the two resulting equations.

Moreover, due to the symmetries
\begin{equation}{\label{Symmetries}}
    G =G^*,\quad \tilde{F} = F^*,
\end{equation}
Equation (\ref{CCUsadel}) is just the complex conjugate of Eq.\ (\ref{Usadel}).
It is therefore sufficient to consider only Eqs.\ (\ref{Usadel}) and (\ref{NC}).

The normalization condition (\ref{NC}) can be resolved with the help of the angular parametrization \cite{Stoof1996.PhysRevB.53.14496,Belzig1999}
\begin{equation}{\label{Angular parameterisation}}
    G=\cos\theta,\quad F= e^{i\chi}\sin\theta,\quad\tilde{F}=e^{-i\chi}\sin\theta.
\end{equation}
In this parametrization, the Usadel equation (\ref{Usadel}) becomes Eqs.\ (\ref{Diffusion})--(\ref{Continuity}).
Also by direct substitution one can get the KL boundary conditions \cite{KL1987,Belzig1999} in the form of Eqs.\ (\ref{Theta boundary})--(\ref{Velocity boundary}) from the expression for functions $F$ and $G$
\begin{equation}{\label{FG boundary conditions}}
    \frac{\sigma_{l}}{g_N}G_{l}^{2}\nabla_{\mathbf{n}}\frac{F_{l}}{G_{l}}= \frac{\sigma_{r}}{g_N}G_{r}^{2}\nabla_{\mathbf{n}}\frac{F_{r}}{G_{r}}=F_{r}G_{l}-F_{l}G_{r},
\end{equation}
where the indices $l$ and $r$ denote the left and right sides of the interface, respectively, and $\mathbf{n}$ is the unit vector perpendicular to the interface.

\section{Order parameter and current}
\label{Appendix B}

\subsection{Arbitrary temperature}

In this Appendix, we present detailed derivation of Eqs.\ (\ref{Delta_1}) and (\ref{Current answer}).

We start with the order parameter. In order to obtain Eq.\ (\ref{Theta-Delta connection}), one has to solve Eq.\ (\ref{Diffusion linearized}) for $\theta_1$ with the boundary condition (\ref{Theta boundary linearized}), which can be included into Eq.\ (\ref{Diffusion linearized}) by employing the Dirac delta function. The result reads
\begin{equation}{\label{DiffusionDirac}}
\frac{d^{2}\theta_{1}}{d z^{2}}+\frac{\Delta_{1}(z)}{\Delta_{0}}\cos\theta_{0}-\frac{\theta_{1}}{\sin\theta_{0}}=\sin2\theta_{0}\left(1-\cos\delta\varphi\right)\delta(z).
\end{equation}
The Fourier transformation of Eq.\ (\ref{DiffusionDirac}) leads to Eq.\ (\ref{Theta-Delta connection}). Substituting Eq.\ (\ref{Theta-Delta connection}) into Eq.\ (\ref{NoLogSelfConsist}), we obtain \begin{multline}{\label{DeltaSums}}
    \frac{\Delta_{1}(k)}{\Delta_{0}}\left( k^{2}\sum_{\omega}\frac{\sin^{2}\theta_{0}}{k^{2}\sin\theta_{0}+1}+\sum_{\omega}\frac{\sin^{3}\theta_{0}}{k^{2}\sin\theta_{0}+1}\right) \\=-2\left(1-\cos\delta\varphi\right) \sum_{\omega}\frac{\sin^{2}\theta_{0}\cos^{2}\theta_{0}}{k^{2}\sin\theta_{0}+1}.
\end{multline}

Since all the sums in this equation converge, we can extend the limits of summation to infinity, formally putting $\omega_D=\infty$.
Using the definition of Eq.\ (\ref{Sums}), we can finally rewrite Eq.\ (\ref{DeltaSums}) in the form of Eq.\ (\ref{Delta_1}).

In order to find the current $J$, we expand Eq.\ (\ref{Current relation}) up to $\alpha^2$, obtaining
\begin{multline}{\label{Current expansion}}
    J=2\pi e \nu_{0}DTS \biggl( \alpha\sum_{\omega}\sin^{2}\theta_{0}\frac{d\chi_{1}}{dx} +\alpha^{2}\sum_{\omega}\sin^{2}\theta_{0}\frac{d\chi_{2}}{dx}
    \\
    +\alpha^{2}\sum_{\omega}2\theta_{1}\sin\theta_{0}\cos\theta_{0}\frac{d\chi_{1}}{dx} \biggr).
\end{multline}
Due to the current conservation, we can calculate the current at any point. Doing this at the interface with the help of the boundary conditions (\ref{Velocity boundary linearized}) and (\ref{Second order velocity boundary}), we obtain
\begin{multline}{\label{CurrentAppendix}}
    J=4\pi e T\nu_{0}DS \frac{g_N}{\sigma}\sin\delta\varphi\sum_{\omega>0}\sin^{2}\theta_{0}
    \\
    \times\left(1+\alpha\frac{\sum_{\omega>0}2\theta_{1}(z=0)\sin\theta_{0}\cos\theta_{0}}{\sum_{\omega>0}\sin^{2}\theta_{0}}\right).
\end{multline}
Here we use the relations $2 e^2\nu_{0}D=\sigma$ and
\begin{equation}\label{relation}
    \frac{2\pi T}{\Delta_{0}}\sum_{\omega>0}\frac{\Delta_{0}^{2}}{\omega^{2}+\Delta_{0}^{2}}=\frac{\pi}{2}\tanh\left(\frac{\Delta_{0}}{2T}\right).
\end{equation}
The interface value $\theta_1(z = 0) $ is calculated using the inverse Fourier transformation of $\theta_1(k)$ given by Eq.\ (\ref{Theta-Delta connection}). As a result,
\begin{multline}{\label{Theta1 at boundary}}
    \theta_{1}(z=0)=\int\limits_{-\infty}^{\infty}\frac{d k}{2\pi}\biggl\{\frac{\sin\theta_{0}\cos\theta_{0}}{k^{2}\sin\theta_{0}+1}
    \\
    \times\left[\frac{\Delta_{1}(k)}{\Delta_0}-2(1-\cos\delta\varphi)\sin\theta_{0}\right]\biggr\}.
\end{multline}
Substituting this into Eq.\ (\ref{CurrentAppendix}), we can write the result for the current in the form of Eqs.\ (\ref{Current answer})--(\ref{V(T)}).

\subsection{Solvable temperature limits}

In this Appendix, we evaluate the sums defined by Eq.\ (\ref{Sums}) in the two limiting cases, $T\to  0$ and $T\to  T_c$.
Then we discuss the corresponding limiting results for the correction $\Delta_{1}(k)$ to the order parameter and for $V(T)$, which determines the corrections to the Josephson current, see Eqs.\ (\ref{Delta_1}) and (\ref{V(T)}).

\subsubsection{\texorpdfstring{$T\to 0$}{T->0}}

In the $T\to 0$ limit, the sums of the form Eq.\ (\ref{Sums}) can be replaced by the integrals
\begin{equation}\label{SumToIntegral}
    L_{n}(k,0)=\int_{0}^{\infty}\frac{1}{\left(w^{2}+1\right)^{\frac{n-1}{2}}}\frac{1}{k^{2}+\sqrt{w^{2}+1}}dw.
\end{equation}
To calculate the sums at an arbitrary $n$, we use the following recurrence identity:
\begin{equation}\label{SumToPropogator}
    L_{n+1}(k,T)=\sum_{j=0}^{n-2}\frac{(-1)^{j}}{k^{2j+2}}L_{n-j}(0,T)-\frac{(-1)^n}{k^{2n-2}}L_{2}(k,T).
\end{equation}
In the $T\to 0$ limit,
\begin{equation}\label{ZeroTemperatureConst}
    L_{n}(0,0)=\frac{\sqrt{\pi}\Gamma\left(\frac{n-1}{2}\right)}{2\Gamma\left(\frac{n}{2}\right)}.
\end{equation}
The sum $L_{2}(k,0)$ has the form
\begin{equation}\label{L2}
    L_{2}(k,0)=\left\{
    \begin{aligned}
    &\frac{1}{\sqrt{1-k^{4}}}\left[\frac{\pi}{2}-\arctan\left(\frac{k^{2}}{\sqrt{1-k^{4}}}\right)\right],\!\!\! & |k|<1,\\
    &\frac{1}{2\sqrt{k^{4}-1}}\ln\left|\frac{\sqrt{k^{4}-1}+k^{2}}{\sqrt{k^{4}-1}-k^{2}}\right|, & |k|>1.
    \end{aligned}
    \right.
\end{equation}
Then, we obtain
\begin{align}
    L_{3}(k,0)&=\frac{\pi}{2k^{2}}-\frac{L_{2}(k,0)}{k^{2}}\label{L3},\\
    L_{4}(k,0)&=\frac{1}{k^{2}}-\frac{\pi}{2k^{4}}+\frac{L_{2}(k,0)}{k^{4}}\label{L4},\\
    L_{5}(k,0)&=\frac{\pi}{4k^{2}}-\frac{1}{k^{4}}+\frac{\pi}{2k^{6}}-\frac{L_{2}(k,0)}{k^{6}}.\label{L5}
\end{align}
Plugging the obtained expressions into Eq.\ (\ref{Delta_1}), we find $\Delta_{1}(k)$.

At the same time, the current (\ref{Current answer}) in determined by $V(T)$ containing an integral with the $L_n$ sums, see Eq.\ (\ref{V(T)}).
Although we are not able to calculate the integral in Eq.\ (\ref{V(T)}) at $T=0$ analytically,  we can do it numerically obtaining $V(0)\approx 0.272$.

\subsubsection{\texorpdfstring{$T\to  T_{c}$}{T->Tc}}

In this limit, $\Delta_0$ becomes small and has the form \cite{Abrikosov1988}
\begin{align}\label{GLDelta}
    \Delta_{0}(T)=\sqrt{8\pi^{2} T_{c}\left(T_{c}-T\right) / 7\zeta\left(3\right)}.
\end{align}
Therefore, we keep only the leading orders in $\Delta_0$ in Eqs.\ (\ref{Delta_1}) and (\ref{V(T)}). Thus, we obtain
\begin{gather}
    \frac{\Delta_{1}(k)}{\Delta_{0}}=-2(1-\cos\delta\varphi)\frac{1}{k^{2}+L_{3}(k,T)/L_{2}(k,T)}\label{Delta_1_GL},\\
    V(T)=\frac{2T_{c}}{\pi^{2}\Delta_{0}}\int\limits_{-\infty}^{\infty}dk\frac{L_{2}(k,T)}{k^{2}+L_{3}(k,T)/L_{2}(k,T)}\label{V(T)GL}.
\end{gather}
In these formulas, we may neglect the $k$ dependence in the $L_n$ sums putting $k=0$.
Indeed, in the $T\to  T_{c}$ limit, we have $\omega\sim T_c$ and $\Delta_0\ll T_c$, hence $\sin \theta_0 \approx \Delta_0/\omega\ll 1$ and $L_3/L_2\sim\Delta_0/T_c\ll1$. From Eq.\ (\ref{Delta_1_GL}) we see that $\Delta_1(k)$ varies on the scale of $k \sim \sqrt{\Delta_0/T_c}$. At the same time, the $L_n$ sums, Eq.\ (\ref{Sums}), vary on the scale of $k \sim \sqrt{T_c/\Delta_0}\gg\sqrt{\Delta_0/T_c}$. We thus obtain
\begin{gather}
    \frac{\Delta_{1}(k)}{\Delta_{0}}=-\frac{2\left(1-\cos\delta\varphi\right)}{k^{2}+7\zeta(3)\Delta_{0}/\pi^{3}T_{c}},
    \label{Delta1(k)} \\
    V(T)=\sqrt{\pi^{3}T_{c} / 28\zeta(3)\Delta_{0}}.
\end{gather}
Substituting Eq.\ (\ref{GLDelta}) into the latter expression, we obtain Eq.\ (\ref{V(T)Answer}).

\section{Second-order perturbation theory for the phases}
\label{Appendix C}

In this Appendix, we calculate the second-order corrections for the phases, $\chi_2(z,\omega)$ and $\varphi_2(z)$.

\subsection{\texorpdfstring{Calculation of $\chi_2$ and $\varphi_2$}{Calculation of chi2 and varphi2}}

We start from finding the auxiliary quantities $\chi_{2}^{(0)}(z,\omega)$ and $\varphi_{2}^{(0)}(z)$. By definition, these functions are continuous at any $z$ and vanish in the bulk. The equations for these functions have the form
\begin{gather}
    \sin\theta_{0}\frac{d^{2}\chi_{2}^{(0)}}{dz^{2}}+2\frac{d\theta_{1}}{dz}\cos\theta_{0}\sin\delta\varphi=\chi_{2}^{(0)}-\varphi_{2}^{(0)},\label{Chi20}\\
    \sum_{|\omega|<\omega_{D}} \left(\chi_{2}^{(0)}(z,\omega)-\varphi_{2}^{(0)}(z)\right)\sin\theta_{0}=0.\label{SelfConsistency20}
\end{gather}
Fourier transformation of Eq.\ (\ref{Chi20}) gives
\begin{equation}
   \chi_{2}^{(0)}(k)= \frac{\varphi_{2}^{(0)}(k)+2ik\theta_{1}(k)\cos\theta_{0}\sin\delta\varphi}{1+k^{2}\sin\theta_{0}} \label{Chi20Connection}
\end{equation}
(we omit the $\omega$ argument of $\chi_2$, $\chi_{2}^{(0)}$, $\theta_1$, and $\theta_0$ for brevity). Substituting Eq.\ (\ref{Chi20Connection}) into Eq.\ (\ref{SelfConsistency20}), we find
\begin{equation}
    \varphi_{2}^{(0)}(k)=\frac{4\pi iT}{\Delta_{0}}\frac{1}{kL_{2}(k)}\sum_{\omega>0}\frac{\theta_{1}(k)\sin\theta_{0}\cos\theta_{0}}{1+k^{2}\sin\theta_{0}}\sin\delta\varphi.\label{Varphi20}
\end{equation}

Our actual problem for finding $\chi_2$ and $\varphi_2$, defined by Eqs.\ (\ref{ContinuityDoubleLinearized})--(\ref{VelocityBoundary2}), is more complicated than the one for  $\chi_2^{(0)}$ and $\varphi_2^{(0)}$ due to two circumstances. First, current conservation leads to nonzero correction to the velocity of the Cooper pairs in the bulk, i.e., $d\varphi_2/dz\neq 0$ at $z\to \infty$, which leads to delta-functional singularity in the Fourier transform of $\chi_2$ and $\varphi_2$. Since $\chi=\varphi$ in the bulk, it is possible to solve the system of equations for $\phi_2=\chi_2-\varphi_2$ from which the singularity drops out. Second, $\chi_2$ can be discontinuous at $z=0$, which leads to a singularity in Eq.\ (\ref{ContinuityDoubleLinearized}),
\begin{multline}
    \sin\theta_{0}\frac{d^{2}\phi_{2}}{dz^{2}}+\sin\theta_{0}\frac{d^{2}\varphi_{2}}{dz^{2}}+2\frac{d\theta_{1}}{dz}\cos\theta_{0}\sin\delta\varphi
    \\
    =\phi_{2}+\delta\chi_{2}\delta^{\prime}(z)\sin\theta_{0}.\label{ContinuityDeltaFunction}
\end{multline}
The Fourier transformation of Eq.\ (\ref{ContinuityDeltaFunction}) gives Eq.\ (\ref{ContinuityDoubleFourier}).

Finally, due to discontinuity at $z=0$, the derivative of $\chi_2$ contains the delta-functional contribution $\delta\chi_2\delta(z)$. The boundary condition (\ref{VelocityBoundary2}) contains only one-sided limits at $z=0$, so in order to write Eq.\ (\ref{VelocityBoundary2}) in the Fourier space, we have to subtract from the Fourier transform the singularity due to the phase jump $\delta\chi_2\delta(z)$.

From the self-consistency equation (\ref{SelfConsistencyDoubleLFourier}) we find the connection between $\varphi_{2}(k)$ and $\delta\chi_2$,
\begin{multline}
    k^{2}\varphi_{2}=\frac{4\pi Tik}{\Delta_{0}L_{2}(k)}\sum_{\omega>0}\frac{\theta_{1}\sin\theta_{0}\cos\theta_{0}\sin\delta\varphi}{1+k^{2}\sin\theta_{0}}\\
    -\frac{2\pi Tik}{\Delta_{0}L_{2}(k)}\sum_{\omega>0}\frac{\delta\chi_{2}\sin^{2}\theta_{0}}{1+k^{2}\sin\theta_{0}}\label{Varphi2ConnectionDeltaChi2}.
\end{multline}
Using Eqs.\ (\ref{ContinuityDoubleFourier}), (\ref{Chi20Connection}), (\ref{Varphi20}), and the definition of the phase functional (\ref{PhaseFunctional}), we obtain
\begin{multline} \label{Phi2(k)}
    \phi_{2}(k)=\chi_{2}^{(0)}(k)-\varphi_{2}^{(0)}(k)\\+\frac{ik\sin\theta_{0}}{1+k^{2}\sin\theta_{0}}\frac{\Phi(k)}{L_{2}(k)}-\frac{ik\delta\chi_{2}\sin\theta_{0}}{1+k^{2}\sin\theta_{0}}.
\end{multline}

The next step is to use the boundary condition (\ref{PhasesFourierBoundary}). Substituting there Eq.\ (\ref{Phi2(k)}), we obtain
\begin{equation}
    \frac{\delta\chi_{2}}{2\sqrt{\sin\theta_{0}}}
    =\int\limits_{-\infty}^{\infty}\frac{dk}{2\pi}\left[ik\left(\chi_{2}^{(0)}-\varphi_{2}^{(0)}\right)-\frac{k^{2}\sin\theta_{0}}{1+k^{2}\sin\theta_{0}}\frac{\Phi}{L_{2}}\right].
\end{equation}
This can be transformed as
\begin{multline}
    \frac{\delta\chi_{2}}{2\sqrt{\sin\theta_{0}}}
    =\frac{d\chi_{2}^{(0)}(z=0)}{dz}-\frac{d\varphi_{2}^{(0)}(z=0)}{dz}
    \\
    +\int\limits_{-\infty}^{\infty}\frac{dk}{2\pi}\frac{1}{1+k^{2}\sin\theta_{0}}\frac{\Phi(k)}{L_2(k)}-\int\limits_{-\infty}^{\infty}\frac{dk}{2\pi}\frac{\Phi(k)}{L_2(k)}.\label{IntEq2}
\end{multline}
Below for brevity we denote $d\chi_{2}^{(0)}(z=0)/dz$ by $\chi_{2}^{\prime(0)}(0)$, and similar notation is used for $\varphi_{2}^{(0)}$.

Now, we multiply Eq.\ (\ref{IntEq2}) by $2\pi T \sin^2\theta_0/\Delta_0$ and sum over $\omega>0$. Then by definition of $L_2$, see Eq.\ (\ref{Sums}), we have
\begin{multline}
    \frac{\pi T}{\Delta_{0}}\sum_{\omega>0}\delta\chi_{2}\sin^{3/2}\theta_{0}-\int\limits_{-\infty}^{\infty}\frac{dk}{2\pi}\Phi(k)
    \\
    =\frac{2\pi T}{\Delta_{0}}\sum_{\omega>0}\sin^{2}\theta_{0} \biggl( \chi_{2}^{\prime(0)}(0)-\varphi_{2}^{\prime(0)}(0)-\int\limits_{-\infty}^{\infty}\frac{dk}{2\pi}\frac{\Phi(k)}{L_{2}(k)} \biggr).
\end{multline}
The left-hand side turns to zero after integration of the phase functional, Eq.\ (\ref{PhaseFunctional}), over $k$. At the same time, in the right-hand side we have a contribution
\begin{equation}
    \int\limits_{-\infty}^{\infty}\frac{dk}{2\pi}\frac{\Phi(k)}{L_{2}(k)}
    =\frac{\sum_{\omega>0}\chi_{2}^{\prime(0)}(0)\sin^{2}\theta_{0}}{\sum_{\omega>0}\sin^{2}\theta_{0}}-\varphi_{2}^{\prime(0)}(0).\label{FInt}
\end{equation}
We denote
\begin{equation}
    \frac{\sum_{\omega>0}\chi_{2}^{\prime(0)}(0)\sin^{2}\theta_{0}}{\sum_{\omega>0}\sin^{2}\theta_{0}}\equiv V_{0}.
\end{equation}
Substituting this result into Eq.\ (\ref{IntEq2}), we obtain
\begin{equation}
    \frac{\delta\chi_{2}}{2\sqrt{\sin\theta_{0}}}=\chi_{2}^{\prime(0)}(0)-V_{0}+\int\limits_{-\infty}^{\infty}\frac{dk}{2\pi}\frac{1}{1+k^{2}\sin\theta_{0}}\frac{\Phi(k)}{L_{2}(k)}.
\end{equation}
In order to calculate $V_0$, we consider the Fourier transform of $\varphi^{\prime}_{2}(z)$, and employing Eq.\ (\ref{Varphi2ConnectionDeltaChi2}), we get
\begin{equation}
    ik\varphi_{2}(k)=ik\varphi_{2}^{(0)}(k)+ \Phi(k) /L_{2}(k) +\beta\delta(k),\label{Phi2Velocity}
\end{equation}
where $\beta$ is an unknown coefficient. Since $\varphi_2(z)$ is a continuous function at $z=0$ and $\varphi^{\prime}_{2}(0)=0$ due to the boundary condition (\ref{VelocityBoundary2}), we obtain
\begin{equation}
    0=\varphi_{2}^{\prime(0)}(0)+\int\limits_{-\infty}^{\infty}\frac{dk}{2\pi}\frac{\Phi(k)}{L_{2}(k)}+\frac{\beta}{2\pi}.
\end{equation}
From Eq.\ (\ref{FInt}), we find
\begin{equation}
    \beta=-2\pi V_{0}.
\end{equation}

Finally, due to the current conservation, we can consider the current in the bulk where $\theta_1=0$. Employing Eq.\ (\ref{Current answer}), we obtain
\begin{multline}
    J_{0}\sin\delta\varphi\left[1-4\alpha(1-\cos\delta\varphi)V\right]\\=J_{0}\left(\sin\delta\varphi+\alpha\frac{d\varphi_{2}(z=\infty)}{dz}\right).
\end{multline}
Expressing $d\varphi_2(z=\infty)/dz$ with the help of Eq.\ (\ref{Phi2Velocity}), we obtain
\begin{equation}
    V_{0}=-\frac{d\varphi_2(z=\infty)}{dz}=4V(1-\cos\delta\varphi)\sin\delta\varphi.
\end{equation}

The answer for the phases $\chi_2$ and $\varphi_2$ thus reads
\begin{align}
    ik\chi_{2}(k)
    &=ik\chi_{2}^{(0)}(k)+\frac{1}{k^{2}\sin\theta_{0}+1}\frac{\Phi(k)}{L_2(k)}
    \notag \\
    &\hphantom{=}-2\pi V_{0} \delta(k)+\frac{\delta\chi_{2}k^{2}\sin\theta_{0}}{k^{2}\sin\theta_{0}+1},\label{chi_2Answer}
    \\
    ik\varphi_{2}(k)&=ik\varphi_{2}^{(0)}(k)+ \Phi(k) / L_2(k) -2\pi V_{0} \delta(k).\label{varphi_2Answer}
\end{align}
The inverse Fourier transformation gives the derivatives $d\chi_2/dz$ and $d\varphi_2/dz$, from which we can find $\chi_2(z,\omega)$ and $\varphi_2(z)$, respectively.

\subsection{Bulk behavior}

From Eqs.\ (\ref{chi_2Answer}) and (\ref{varphi_2Answer}), one can see that in the bulk, the phases $\chi_2$ and $\varphi_2$ are equal and vary linearly as
\begin{equation}
	\chi_2(z\to \infty,
	\omega)=a\sgn z+bz,
\end{equation}
with constant coefficients $a$ and $b$. Our goal now is to find them.

Since $\chi_2(k)$ is an odd function, we can write
\begin{equation}
	\chi_2(z)=\int\limits_{-\infty}^{\infty}\frac{dk}{2\pi}\frac{\sin{kz}}{k}ik\chi_2(k).\label{NewtonForChi2}
\end{equation}
Substituting here the relation
\begin{equation}
	\sin (kz) / \pi k \xrightarrow[z\to\infty]{} \sgn(z)\delta(k) \label{DeltaFunctionRelation}
\end{equation}
and employing Eq.\ (\ref{chi_2Answer}), we find
\begin{equation}
 a=\frac{1}{2}\lim_{k\to 0} \left(ik\chi_{2}^{(0)}(k)+\frac{\Phi(0)}{L_2(0)}\right),
 \quad
 b=-V_0.\\
\end{equation}
The constant $a$ can be found with the use of Eq.\ (\ref{Chi20Connection}), and in terms of the sums defined in Eq.\ (\ref{Sums}), it acquires the form
\begin{multline}
	a=2(1-\cos\delta\varphi)\sin\delta \varphi
    \\
    \times\left[\frac{(L_2-L_4)^2+L_3(L_3-L_5)}{L_2L_3}+\frac{\Phi}{L_2}\right]_{k=0}.
\end{multline}
Both constants $a$ and $b$ do not depend on $\omega$, which shows that $\chi_2=\varphi_2$ in the bulk. Unlike the constant $b=-V_0$, the constant $a$ at arbitrary temperatures cannot be found without solving the integral equation (\ref{PhaseShiftEquation}), since it is determined by the phase functional $\Phi(k)$. At the same time, in the limit $T=0$,  its value can be found numerically and equals $a\approx 0.544(1-\cos\delta\varphi)\sin\delta\varphi$.

\bibliography{bib}

\begin{thebibliography}{28}%
\makeatletter
\providecommand \@ifxundefined [1]{%
 \@ifx{#1\undefined}
}%
\providecommand \@ifnum [1]{%
 \ifnum #1\expandafter \@firstoftwo
 \else \expandafter \@secondoftwo
 \fi
}%
\providecommand \@ifx [1]{%
 \ifx #1\expandafter \@firstoftwo
 \else \expandafter \@secondoftwo
 \fi
}%
\providecommand \natexlab [1]{#1}%
\providecommand \enquote  [1]{``#1''}%
\providecommand \bibnamefont  [1]{#1}%
\providecommand \bibfnamefont [1]{#1}%
\providecommand \citenamefont [1]{#1}%
\providecommand \href@noop [0]{\@secondoftwo}%
\providecommand \href [0]{\begingroup \@sanitize@url \@href}%
\providecommand \@href[1]{\@@startlink{#1}\@@href}%
\providecommand \@@href[1]{\endgroup#1\@@endlink}%
\providecommand \@sanitize@url [0]{\catcode `\\12\catcode `\$12\catcode
  `\&12\catcode `\#12\catcode `\^12\catcode `\_12\catcode `\%12\relax}%
\providecommand \@@startlink[1]{}%
\providecommand \@@endlink[0]{}%
\providecommand \url  [0]{\begingroup\@sanitize@url \@url }%
\providecommand \@url [1]{\endgroup\@href {#1}{\urlprefix }}%
\providecommand \urlprefix  [0]{URL }%
\providecommand \Eprint [0]{\href }%
\providecommand \doibase [0]{https://doi.org/}%
\providecommand \selectlanguage [0]{\@gobble}%
\providecommand \bibinfo  [0]{\@secondoftwo}%
\providecommand \bibfield  [0]{\@secondoftwo}%
\providecommand \translation [1]{[#1]}%
\providecommand \BibitemOpen [0]{}%
\providecommand \bibitemStop [0]{}%
\providecommand \bibitemNoStop [0]{.\EOS\space}%
\providecommand \EOS [0]{\spacefactor3000\relax}%
\providecommand \BibitemShut  [1]{\csname bibitem#1\endcsname}%
\let\auto@bib@innerbib\@empty
\bibitem [{\citenamefont {Tinkham}(2004)}]{TinkhamBook}%
  \BibitemOpen
  \bibfield  {author} {\bibinfo {author} {\bibfnamefont {M.}~\bibnamefont
  {Tinkham}},\ }\href@noop {} {\emph {\bibinfo {title} {Introduction to
  Superconductivity (2nd edition)}}}\ (\bibinfo  {publisher} {Dover},\ \bibinfo
  {address} {New York},\ \bibinfo {year} {2004})\BibitemShut {NoStop}%
\bibitem [{\citenamefont {Abrikosov}\ \emph {et~al.}(1977)\citenamefont
  {Abrikosov}, \citenamefont {Gor'kov},\ and\ \citenamefont
  {Dzyaloshinski}}]{AGDBookEng}%
  \BibitemOpen
  \bibfield  {author} {\bibinfo {author} {\bibfnamefont {A.~A.}\ \bibnamefont
  {Abrikosov}}, \bibinfo {author} {\bibfnamefont {L.~P.}\ \bibnamefont
  {Gor'kov}},\ and\ \bibinfo {author} {\bibfnamefont {I.~E.}\ \bibnamefont
  {Dzyaloshinski}},\ }\href@noop {} {\emph {\bibinfo {title} {Methods of
  Quantum Field Theory in Statistical Physics}}}\ (\bibinfo  {publisher}
  {Dover},\ \bibinfo {address} {New York},\ \bibinfo {year} {1977})\BibitemShut
  {NoStop}%
\bibitem [{\citenamefont {Zaikin}\ and\ \citenamefont
  {Zharkov}(1981)}]{Zaikin1981Eng}%
  \BibitemOpen
  \bibfield  {author} {\bibinfo {author} {\bibfnamefont {A.~D.}\ \bibnamefont
  {Zaikin}}\ and\ \bibinfo {author} {\bibfnamefont {G.~F.}\ \bibnamefont
  {Zharkov}},\ }\bibfield  {title} {\bibinfo {title} {Theory of wide dirty
  \textit{SNS} junctions},\ }\href@noop {} {\bibfield  {journal} {\bibinfo
  {journal} {Sov. J. Low Temp. Phys.}\ }\textbf {\bibinfo {volume} {7}},\
  \bibinfo {pages} {184} (\bibinfo {year} {1981})},\ \bibinfo {note} {[Fiz.
  Nizk. Temp., \textbf{7}, 375 (1981)]}\BibitemShut {NoStop}%
\bibitem [{\citenamefont {Stoof}\ and\ \citenamefont
  {Nazarov}(1996)}]{Stoof1996.PhysRevB.53.14496}%
  \BibitemOpen
  \bibfield  {author} {\bibinfo {author} {\bibfnamefont {T.~H.}\ \bibnamefont
  {Stoof}}\ and\ \bibinfo {author} {\bibfnamefont {{\relax Yu}.~V.}\
  \bibnamefont {Nazarov}},\ }\bibfield  {title} {\bibinfo {title}
  {Kinetic-equation approach to diffusive superconducting hybrid devices},\
  }\href {https://doi.org/10.1103/PhysRevB.53.14496} {\bibfield  {journal}
  {\bibinfo  {journal} {Phys. Rev. B}\ }\textbf {\bibinfo {volume} {53}},\
  \bibinfo {pages} {14496} (\bibinfo {year} {1996})}\BibitemShut {NoStop}%
\bibitem [{\citenamefont {Belzig}\ \emph {et~al.}(1999)\citenamefont {Belzig},
  \citenamefont {Wilhelm}, \citenamefont {Bruder}, \citenamefont {Sch{\"o}n},\
  and\ \citenamefont {Zaikin}}]{Belzig1999}%
  \BibitemOpen
  \bibfield  {author} {\bibinfo {author} {\bibfnamefont {W.}~\bibnamefont
  {Belzig}}, \bibinfo {author} {\bibfnamefont {F.~K.}\ \bibnamefont {Wilhelm}},
  \bibinfo {author} {\bibfnamefont {C.}~\bibnamefont {Bruder}}, \bibinfo
  {author} {\bibfnamefont {G.}~\bibnamefont {Sch{\"o}n}},\ and\ \bibinfo
  {author} {\bibfnamefont {A.~D.}\ \bibnamefont {Zaikin}},\ }\bibfield  {title}
  {\bibinfo {title} {Quasiclassical {Green}’s function approach to mesoscopic
  superconductivity},\ }\href {https://doi.org/10.1006/spmi.1999.0710}
  {\bibfield  {journal} {\bibinfo  {journal} {Superlattices Microstruct.}\
  }\textbf {\bibinfo {volume} {25}},\ \bibinfo {pages} {1251} (\bibinfo {year}
  {1999})}\BibitemShut {NoStop}%
\bibitem [{\citenamefont {Golubov}\ \emph {et~al.}(2002)\citenamefont
  {Golubov}, \citenamefont {Kupriyanov},\ and\ \citenamefont
  {Fominov}}]{Golubov2002Eng}%
  \BibitemOpen
  \bibfield  {author} {\bibinfo {author} {\bibfnamefont {A.~A.}\ \bibnamefont
  {Golubov}}, \bibinfo {author} {\bibfnamefont {M.~{\relax Yu}.}\ \bibnamefont
  {Kupriyanov}},\ and\ \bibinfo {author} {\bibfnamefont {{\relax Ya}.~V.}\
  \bibnamefont {Fominov}},\ }\bibfield  {title} {\bibinfo {title} {Critical
  current in {SFIFS} junctions},\ }\href {https://doi.org/10.1134/1.1475721}
  {\bibfield  {journal} {\bibinfo  {journal} {JETP Lett.}\ }\textbf {\bibinfo
  {volume} {75}},\ \bibinfo {pages} {190} (\bibinfo {year} {2002})},\ \bibinfo
  {note} {[Pis’ma Zh. Eksp. Teor. Fiz., \textbf{75}, 223 (2002)]}\BibitemShut
  {NoStop}%
\bibitem [{\citenamefont {Josephson}(1962)}]{Josephson1962}%
  \BibitemOpen
  \bibfield  {author} {\bibinfo {author} {\bibfnamefont {B.~D.}\ \bibnamefont
  {Josephson}},\ }\bibfield  {title} {\bibinfo {title} {Possible new effects in
  superconductive tunnelling},\ }\href
  {https://doi.org/10.1016/0031-9163(62)91369-0} {\bibfield  {journal}
  {\bibinfo  {journal} {Phys. Letters}\ }\textbf {\bibinfo {volume} {1}},\
  \bibinfo {pages} {251} (\bibinfo {year} {1962})}\BibitemShut {NoStop}%
\bibitem [{\citenamefont {Ambegaokar}\ and\ \citenamefont
  {Baratoff}(1963)}]{Ambegaokar1963}%
  \BibitemOpen
  \bibfield  {author} {\bibinfo {author} {\bibfnamefont {V.}~\bibnamefont
  {Ambegaokar}}\ and\ \bibinfo {author} {\bibfnamefont {A.}~\bibnamefont
  {Baratoff}},\ }\bibfield  {title} {\bibinfo {title} {Tunneling between
  superconductors},\ }\href {https://doi.org/10.1103/PhysRevLett.10.486}
  {\bibfield  {journal} {\bibinfo  {journal} {Phys. Rev. Lett.}\ }\textbf
  {\bibinfo {volume} {10}},\ \bibinfo {pages} {486} (\bibinfo {year}
  {1963})}\BibitemShut {NoStop}%
\bibitem [{\citenamefont {Likharev}(1979)}]{Likharev1979.RevModPhys.51.101}%
  \BibitemOpen
  \bibfield  {author} {\bibinfo {author} {\bibfnamefont {K.~K.}\ \bibnamefont
  {Likharev}},\ }\bibfield  {title} {\bibinfo {title} {Superconducting weak
  links},\ }\href {https://doi.org/10.1103/RevModPhys.51.101} {\bibfield
  {journal} {\bibinfo  {journal} {Rev. Mod. Phys.}\ }\textbf {\bibinfo {volume}
  {51}},\ \bibinfo {pages} {101} (\bibinfo {year} {1979})}\BibitemShut
  {NoStop}%
\bibitem [{\citenamefont {Golubov}\ \emph {et~al.}(2004)\citenamefont
  {Golubov}, \citenamefont {Kupriyanov},\ and\ \citenamefont
  {Il'ichev}}]{Golubov2004review}%
  \BibitemOpen
  \bibfield  {author} {\bibinfo {author} {\bibfnamefont {A.~A.}\ \bibnamefont
  {Golubov}}, \bibinfo {author} {\bibfnamefont {M.~{\relax Yu}.}\ \bibnamefont
  {Kupriyanov}},\ and\ \bibinfo {author} {\bibfnamefont {E.}~\bibnamefont
  {Il'ichev}},\ }\bibfield  {title} {\bibinfo {title} {The current-phase
  relation in {Josephson} junctions},\ }\href
  {https://doi.org/10.1103/RevModPhys.76.411} {\bibfield  {journal} {\bibinfo
  {journal} {Rev. Mod. Phys.}\ }\textbf {\bibinfo {volume} {76}},\ \bibinfo
  {pages} {411} (\bibinfo {year} {2004})}\BibitemShut {NoStop}%
\bibitem [{\citenamefont {Ivanov}\ \emph {et~al.}(1981)\citenamefont {Ivanov},
  \citenamefont {Kupriyanov}, \citenamefont {Likharev}, \citenamefont
  {Meriakri},\ and\ \citenamefont {Snigirev}}]{Ivanov1981Eng}%
  \BibitemOpen
  \bibfield  {author} {\bibinfo {author} {\bibfnamefont {Z.~G.}\ \bibnamefont
  {Ivanov}}, \bibinfo {author} {\bibfnamefont {M.~{\relax Yu}.}\ \bibnamefont
  {Kupriyanov}}, \bibinfo {author} {\bibfnamefont {K.~K.}\ \bibnamefont
  {Likharev}}, \bibinfo {author} {\bibfnamefont {S.~V.}\ \bibnamefont
  {Meriakri}},\ and\ \bibinfo {author} {\bibfnamefont {O.~V.}\ \bibnamefont
  {Snigirev}},\ }\bibfield  {title} {\bibinfo {title} {Boundary conditions for
  the {Usadel} and {Eilenberger} equations, and properties of ``dirty''
  \textit{SNS} sandwich-type junctions},\ }\href@noop {} {\bibfield  {journal}
  {\bibinfo  {journal} {Sov. J. Low Temp. Phys.}\ }\textbf {\bibinfo {volume}
  {7}},\ \bibinfo {pages} {274} (\bibinfo {year} {1981})},\ \bibinfo {note}
  {[Fiz. Nizk. Temp., \textbf{7}, 560 (1981)]}\BibitemShut {NoStop}%
\bibitem [{\citenamefont {Kupriyanov}\ and\ \citenamefont
  {Lukichev}(1982)}]{Kupriyanov1982Eng}%
  \BibitemOpen
  \bibfield  {author} {\bibinfo {author} {\bibfnamefont {M.~{\relax Yu}.}\
  \bibnamefont {Kupriyanov}}\ and\ \bibinfo {author} {\bibfnamefont {V.~F.}\
  \bibnamefont {Lukichev}},\ }\bibfield  {title} {\bibinfo {title} {The
  proximity effect in electrodes and the steady-state properties of {Josephson}
  \textit{SNS} structures},\ }\href@noop {} {\bibfield  {journal} {\bibinfo
  {journal} {Sov. J. Low Temp. Phys.}\ }\textbf {\bibinfo {volume} {8}},\
  \bibinfo {pages} {526} (\bibinfo {year} {1982})},\ \bibinfo {note} {[Fiz.
  Nizk. Temp., \textbf{8}, 1045 (1982)]}\BibitemShut {NoStop}%
\bibitem [{\citenamefont {Zubkov}\ and\ \citenamefont
  {Kupriyanov}(1983)}]{Zubkov1983Eng}%
  \BibitemOpen
  \bibfield  {author} {\bibinfo {author} {\bibfnamefont {A.~A.}\ \bibnamefont
  {Zubkov}}\ and\ \bibinfo {author} {\bibfnamefont {M.~{\relax Yu}.}\
  \bibnamefont {Kupriyanov}},\ }\bibfield  {title} {\bibinfo {title} {Influence
  of depairing in electrodes on the steady-state properties of weak links},\
  }\href@noop {} {\bibfield  {journal} {\bibinfo  {journal} {Sov. J. Low Temp.
  Phys.}\ }\textbf {\bibinfo {volume} {9}},\ \bibinfo {pages} {279} (\bibinfo
  {year} {1983})},\ \bibinfo {note} {[Fiz. Nizk. Temp., \textbf{9}, 548
  (1983)]}\BibitemShut {NoStop}%
\bibitem [{\citenamefont {Barash}(2012)}]{Barash2012.PhysRevB.85.100503}%
  \BibitemOpen
  \bibfield  {author} {\bibinfo {author} {\bibfnamefont {{\relax Yu}.~S.}\
  \bibnamefont {Barash}},\ }\bibfield  {title} {\bibinfo {title} {Anharmonic
  {Josephson} current in junctions with an interface pair breaking},\ }\href
  {https://doi.org/10.1103/PhysRevB.85.100503} {\bibfield  {journal} {\bibinfo
  {journal} {Phys. Rev. B}\ }\textbf {\bibinfo {volume} {85}},\ \bibinfo
  {pages} {100503(R)} (\bibinfo {year} {2012})}\BibitemShut {NoStop}%
\bibitem [{\citenamefont {Barash}(2014)}]{Barash2014Eng}%
  \BibitemOpen
  \bibfield  {author} {\bibinfo {author} {\bibfnamefont {{\relax Yu}.~S.}\
  \bibnamefont {Barash}},\ }\bibfield  {title} {\bibinfo {title} {Interfacial
  pair breaking and planar weak links with an anharmonic current--phase
  relation},\ }\href {https://doi.org/10.1134/S0021364014150041} {\bibfield
  {journal} {\bibinfo  {journal} {JETP Lett.}\ }\textbf {\bibinfo {volume}
  {100}},\ \bibinfo {pages} {205} (\bibinfo {year} {2014})},\ \bibinfo {note}
  {[Pis’ma Zh. Eksp. Teor. Fiz., \textbf{100}, 226 (2014)]}\BibitemShut
  {NoStop}%
\bibitem [{\citenamefont {Sols}\ and\ \citenamefont
  {Ferrer}(1994)}]{Sols1994.PhysRevB.49.15913}%
  \BibitemOpen
  \bibfield  {author} {\bibinfo {author} {\bibfnamefont {F.}~\bibnamefont
  {Sols}}\ and\ \bibinfo {author} {\bibfnamefont {J.}~\bibnamefont {Ferrer}},\
  }\bibfield  {title} {\bibinfo {title} {Crossover from the {Josephson} effect
  to bulk superconducting flow},\ }\href
  {https://doi.org/10.1103/PhysRevB.49.15913} {\bibfield  {journal} {\bibinfo
  {journal} {Phys. Rev. B}\ }\textbf {\bibinfo {volume} {49}},\ \bibinfo
  {pages} {15913} (\bibinfo {year} {1994})}\BibitemShut {NoStop}%
\bibitem [{\citenamefont {Pastukh}\ \emph {et~al.}(2017)\citenamefont
  {Pastukh}, \citenamefont {Shutovskii},\ and\ \citenamefont
  {Sakhnyuk}}]{Pastukh2017}%
  \BibitemOpen
  \bibfield  {author} {\bibinfo {author} {\bibfnamefont {O.~{\relax Yu}.}\
  \bibnamefont {Pastukh}}, \bibinfo {author} {\bibfnamefont {A.~M.}\
  \bibnamefont {Shutovskii}},\ and\ \bibinfo {author} {\bibfnamefont {V.~E.}\
  \bibnamefont {Sakhnyuk}},\ }\bibfield  {title} {\bibinfo {title} {The effect
  of depairing on the current-phase relation in {SIS} junctions in the presence
  of nonmagnetic impurities of arbitrary concentration},\ }\href
  {https://doi.org/10.1063/1.4985972} {\bibfield  {journal} {\bibinfo
  {journal} {Low Temp. Phys.}\ }\textbf {\bibinfo {volume} {43}},\ \bibinfo
  {pages} {664} (\bibinfo {year} {2017})},\ \bibinfo {note} {[Fiz. Nizk. Temp.
  \textbf{43}, 835 (2017)]}\BibitemShut {NoStop}%
\bibitem [{\citenamefont {Kupriyanov}(1992)}]{Kupriyanov1992}%
  \BibitemOpen
  \bibfield  {author} {\bibinfo {author} {\bibfnamefont {M.~{\relax Yu}.}\
  \bibnamefont {Kupriyanov}},\ }\bibfield  {title} {\bibinfo {title} {Effect of
  a finite transmission of the insulating layer on the properties of {SIS}
  tunnel junctions},\ }\href@noop {} {\bibfield  {journal} {\bibinfo  {journal}
  {JETP Lett.}\ }\textbf {\bibinfo {volume} {56}},\ \bibinfo {pages} {399}
  (\bibinfo {year} {1992})},\ \bibinfo {note} {[Pis’ma Zh. Eksp. Teor. Fiz.,
  \textbf{56}, 414 (1992)]}\BibitemShut {NoStop}%
\bibitem [{\citenamefont {Golubov}\ and\ \citenamefont
  {Kupriyanov}(2005)}]{GK2005}%
  \BibitemOpen
  \bibfield  {author} {\bibinfo {author} {\bibfnamefont {A.~A.}\ \bibnamefont
  {Golubov}}\ and\ \bibinfo {author} {\bibfnamefont {M.~{\relax Yu}.}\
  \bibnamefont {Kupriyanov}},\ }\bibfield  {title} {\bibinfo {title} {The
  current phase relation in {Josephson} tunnel junctions},\ }\href
  {https://doi.org/10.1134/1.1944074} {\bibfield  {journal} {\bibinfo
  {journal} {JETP Lett.}\ }\textbf {\bibinfo {volume} {81}},\ \bibinfo {pages}
  {335} (\bibinfo {year} {2005})},\ \bibinfo {note} {[Pis’ma Zh. Eksp. Teor.
  Fiz., \textbf{81}, 419 (2005)]}\BibitemShut {NoStop}%
\bibitem [{\citenamefont {Usadel}(1970)}]{Usadel1970}%
  \BibitemOpen
  \bibfield  {author} {\bibinfo {author} {\bibfnamefont {K.~D.}\ \bibnamefont
  {Usadel}},\ }\bibfield  {title} {\bibinfo {title} {Generalized diffusion
  equation for superconducting alloys},\ }\href
  {https://doi.org/10.1103/PhysRevLett.25.507} {\bibfield  {journal} {\bibinfo
  {journal} {Phys. Rev. Lett.}\ }\textbf {\bibinfo {volume} {25}},\ \bibinfo
  {pages} {507} (\bibinfo {year} {1970})}\BibitemShut {NoStop}%
\bibitem [{\citenamefont {Kuprianov}\ and\ \citenamefont
  {Lukichev}(1987)}]{KL1987}%
  \BibitemOpen
  \bibfield  {author} {\bibinfo {author} {\bibfnamefont {M.~{\relax Yu}.}\
  \bibnamefont {Kuprianov}}\ and\ \bibinfo {author} {\bibfnamefont {V.~F.}\
  \bibnamefont {Lukichev}},\ }\bibfield  {title} {\bibinfo {title} {Influence
  of boundary transparency on the critical current of ``dirty'' {SS'S}
  structures},\ }\href@noop {} {\bibfield  {journal} {\bibinfo  {journal}
  {JETP}\ }\textbf {\bibinfo {volume} {94}},\ \bibinfo {pages} {1163} (\bibinfo
  {year} {1987})},\ \bibinfo {note} {[Zh. Eksp. Teor. Fiz., \textbf{94}, 139
  (1987)]}\BibitemShut {NoStop}%
\bibitem [{\citenamefont {Svidzinsky}(1982)}]{Svidzinsky1982}%
  \BibitemOpen
  \bibfield  {author} {\bibinfo {author} {\bibfnamefont {A.~V.}\ \bibnamefont
  {Svidzinsky}},\ }\href@noop {} {\emph {\bibinfo {title} {{Spatially
  Non-Uniform Problems in the Theory of Superconductivity}}}}\ (\bibinfo
  {publisher} {Nauka, Moscow},\ \bibinfo {year} {1982})\ \bibinfo {note} {[in
  Russian]}\BibitemShut {NoStop}%
\bibitem [{\citenamefont {Furusaki}\ and\ \citenamefont
  {Tsukada}(1991)}]{Furusaki1991}%
  \BibitemOpen
  \bibfield  {author} {\bibinfo {author} {\bibfnamefont {A.}~\bibnamefont
  {Furusaki}}\ and\ \bibinfo {author} {\bibfnamefont {M.}~\bibnamefont
  {Tsukada}},\ }\bibfield  {title} {\bibinfo {title} {{DC} {Josephson} effect
  and {Andreev} reflection},\ }\href
  {https://doi.org/10.1016/0038-1098(91)90201-6} {\bibfield  {journal}
  {\bibinfo  {journal} {Solid State Commun.}\ }\textbf {\bibinfo {volume}
  {78}},\ \bibinfo {pages} {299} (\bibinfo {year} {1991})}\BibitemShut
  {NoStop}%
\bibitem [{\citenamefont {Bagwell}(1994)}]{Bagwell1994.PhysRevB.49.6841}%
  \BibitemOpen
  \bibfield  {author} {\bibinfo {author} {\bibfnamefont {P.~F.}\ \bibnamefont
  {Bagwell}},\ }\bibfield  {title} {\bibinfo {title} {Critical current of a
  one-dimensional superconductor},\ }\href
  {https://doi.org/10.1103/PhysRevB.49.6841} {\bibfield  {journal} {\bibinfo
  {journal} {Phys. Rev. B}\ }\textbf {\bibinfo {volume} {49}},\ \bibinfo
  {pages} {6841} (\bibinfo {year} {1994})}\BibitemShut {NoStop}%
\bibitem [{Note1()}]{Note1}%
  \BibitemOpen
  \bibinfo {note} {In particular, at $T=0$ we have $\xi (0)/\xi _\protect
  \mathrm {GL}(0) = 2 e^{C/2}/\pi \approx 0.85$, where $C\approx 0.577$ is
  Euler's constant}\BibitemShut {NoStop}%
\bibitem [{\citenamefont {Nazarov}(1999)}]{Nazarov1999}%
  \BibitemOpen
  \bibfield  {author} {\bibinfo {author} {\bibfnamefont {{\relax Yu}.~V.}\
  \bibnamefont {Nazarov}},\ }\bibfield  {title} {\bibinfo {title} {Novel
  circuit theory of {Andreev} reflection},\ }\href
  {https://doi.org/10.1006/spmi.1999.0738} {\bibfield  {journal} {\bibinfo
  {journal} {Superlattices Microstruct.}\ }\textbf {\bibinfo {volume} {25}},\
  \bibinfo {pages} {1221} (\bibinfo {year} {1999})}\BibitemShut {NoStop}%
\bibitem [{Note2()}]{Note2}%
  \BibitemOpen
  \bibinfo {note} {In Ref.\ \cite {GK2005}, the perturbation theory was
  developed in the coordinate space. Two issues indicate that the presented
  form of solution is not rigorous (we call it ``conjectured''). (i)~In Eq.\
  (31) of Ref.\ \cite {GK2005}, the order parameter and the quantity
  parameterizing the Green functions are expanded in the system of decaying
  exponents. The system does not form a full basis in the functional space,
  which means that actually only a certain class of functions is considered.
  (ii)~Equation (34) in Ref.\ \cite {GK2005} is obtained from Eqs.\ (32) and
  (33) according to the procedure described below Eq.\ (33). This procedure
  leads to equality between two sums running over \protect \emph {different}
  quantities ($\omega $ and $\Omega $). In order to obtain Eq.\ (34), one
  should equate term-by-term the elements of these \protect \emph {different}
  sums. This assumption also implies a certain conjecture about the form of
  solution.}\BibitemShut {Stop}%
\bibitem [{\citenamefont {Abrikosov}(1988)}]{Abrikosov1988}%
  \BibitemOpen
  \bibfield  {author} {\bibinfo {author} {\bibfnamefont {A.~A.}\ \bibnamefont
  {Abrikosov}},\ }\href@noop {} {\emph {\bibinfo {title} {{Fundamentals of the
  Theory of Metals}}}}\ (\bibinfo  {publisher} {NorthHolland, Amsterdam},\
  \bibinfo {year} {1988})\BibitemShut {NoStop}%
\end{thebibliography}%

\end{document}